\documentclass[11pt]{article}
\pdfoutput=1
\usepackage{jcapmod}
\usepackage{booktabs}
\usepackage[english]{babel}
\usepackage{amsmath,amssymb,amsbsy,amstext, amsthm, simplewick}
\usepackage{graphicx}
\usepackage{amsfonts}
\usepackage{amssymb}
\usepackage{exscale,relsize}
\usepackage[makeroom]{cancel}
\usepackage{soul}
\usepackage{float}
\usepackage[utf8]{inputenc}
\RequirePackage{color}
\usepackage{breqn}

\usepackage{colortbl}
\definecolor{green2}{cmyk}{0, 1, 0.5, 0}
\definecolor{lightgreen}{cmyk}{0.2, 0, 0.2, 0.2}
\definecolor{lightgray}{cmyk}{0.1,0.2,0,0.1}
\definecolor{lightgray2}{cmyk}{0.4,0.4,0,0.8}
\definecolor{black}{cmyk}{1.0,1.0,1.0,1.0}

\allowdisplaybreaks[1]

\usepackage{colortbl}
\definecolor{lightgreen}{cmyk}{0.2, 0, 0.2, 0.2}
\definecolor{lightgray}{cmyk}{0.1,0.2,0,0.1}
\definecolor{lightgray2}{cmyk}{0.1,0.1,0,0.1}

\makeatletter
\newlength{\apb@width}
\newcommand{\autoparbox}[2][c]{\settowidth{\apb@width}{#2}\parbox[#1]{\apb@width}{#2}}

\makeatother

\numberwithin{equation}{section}

\def\beq{\begin{equation}}
\def\eeq{\end{equation}}

\def\bea{\begin{eqnarray}}
\def\eea{\end{eqnarray}}
\def\bse{\begin{dmath}}
\def\ese{\end{dmath}}

\def\d{{\rm d}}

\def\d{{\rm d}}

\def\del{\partial}
\def\Mp{M_{\rm Pl}}

\def\fr{\frac}

\def\g{{\tt g}}
\def\0{{\boldsymbol 0}}

\def\V{{\mathcal{V}}}
\def\taa{{\tilde{a}}}
\def\tA{{\tilde{A}}}

\def\fr{\frac}

\DeclareSymbolFont{extraup}{U}{zavm}{m}{n}
\DeclareMathSymbol{\varheart}{\mathalpha}{extraup}{86}
\DeclareMathSymbol{\vardiamond}{\mathalpha}{extraup}{87}

\DeclareRobustCommand{\SkipTocEntry}[4]{}

\begin{document}

\begin{titlepage}
	
	\setcounter{page}{1} \baselineskip=15.5pt \thispagestyle{empty}
	
	\bigskip\
	
	\vspace{1cm}
	\begin{center}
		
		{\fontsize{20}{28}\selectfont  \sffamily \bfseries 
		On chromonatural inflation in string theory }
		
	\end{center}
	
	\vspace{0.2cm}
	
	\begin{center}
		{\fontsize{13}{30}\selectfont Jonathan Holland$^{}$, Ivonne Zavala$^{}$, 
			Gianmassimo Tasinato$^{}$}
	\end{center}

	\begin{center}
		
		\vskip 8pt
		\textsl{
			Department of Physics, Swansea University, Swansea, SA2 8PP, UK}\\
		\vskip 7pt
		
	\end{center}

	\vspace{1.2cm}
	\hrule \vspace{0.3cm}
	\noindent {\sffamily \bfseries Abstract} \\[0.1cm]
	Sourced gravitational waves in chromonatural inflation (CNI) can give rise to a chiral  spectrum of tensor fluctuations  that is  considerably enhanced relative to the vacuum fluctuations.   If the field content of CNI acts purely as a spectator (SCNI), the inflationary sector can be consistent with current data making SCNI very appealing   in view of  future observations.    
	   We investigate the prospects of embedding SCNI in string theory,  in the  framework of K\"ahler inflation in type IIB large volume string compactifications, with a spectator sector  associated with  gaugino condensation on multiply magnetised D7-branes. 
	We first   introduce a generalised field theory framework that describes non-trivial multifield  inflation coupled to gauge fields of the form  generically arising  in supergravity and string theory.  We then use these results to study numerically and analytically both the background evolution and the dynamics of cosmological perturbations.  
	We show that a successful inflationary background evolution  with a large enhancement of the gravitational wave spectrum and a controllable 
   backreaction from the amplified tensor fluctuations can be achieved by considering suitable values of three parameters present in our scenario: the magnetic flux, the degree of the condensing gauge group and the wrapping number of the D7-brane. 
	On the other hand, the required values for these quantities may present a challenge for its successful realisation within string theory. We also discuss these challenges and  the model  from the viewpoint of the weak gravity conjecture.  
	
	\vskip 10pt
	\hrule
	
	\vspace{0.6cm}
\end{titlepage}

\tableofcontents

\section{Introduction}

Cosmological inflation remains the leading paradigm to explain large scale homogeneity and isotropy of the observable universe as well as the origin of the large scale structure we observe today. In its simplest realisation, this early-universe accelerated expansion is driven by  a scalar field, the inflaton, whose potential energy dominates. 
During inflation, the quantum fluctuations in the inflaton and metric tensor fields were stretched to observables scales, and set up the initial conditions for structure growth.
These fluctuations induce tiny temperature differences in the cosmic microwave background  (CMB), recently observed to high precision by the Planck satellite \cite{Planck18}. 

The mechanism of inflation  makes robust predictions for the produced  primordial inhomogeneities.   Namely,  they  are  adiabatic,  approximately  scale-invariant  and  nearly  Gaussian.   All  these  properties  are  in  good  agreement  with current  observations  \cite{Planck18}.  Inflation also predicts the existence of a primordial gravitational wave (PGW) spectrum, whose amplitude  depends on the inflationary model.  
While  vanilla models of inflation are in  agreement with the most recent data, ongoing and future experiments will allow us to take a step forward in testing the inflationary paradigm. Particularly interesting are the PGWs produced by inflation, which lead to a distinctive B-mode pattern in the CMB polarisation \cite{ZU,KamionB}, that is being searched for by a  range of ground-based, balloon and satellite experiments \cite{SPT3G,simons,cmb-s4,class,Lbird,pico}. 
Current bounds on the tensor-to-scalar ratio $r$ from Planck and BICEP/Keck restrict $r <0.056$ \cite{Planck18}, but future experiments are likely to be able to reach a sensitivity of order $\Delta r\simeq 10^{-3}$. 
If the PGWs are entirely sourced from vacuum fluctuations, $r$ can be directly related to the 
energy scale of inflation, $V^{1/4} \sim 1.8\times 10^{16} {\rm GeV} (r/0.1)^{1/4}$ and the  inflaton field displacement during inflation, $\Delta\phi\gtrsim \times (r/0.002)^{1/2}\Mp$ via the Lyth bound \cite{Lyth,Lotfi} (taking into account that $r$ does not remain constant, the bound is much stronger  \cite{Garcia-Bellido:2014wfa}).
 Thus a clear understanding of the  B-mode measurements is tightly linked  to a correct interpretation of their source. 

In  theories beyond the standard model of particle physics and cosmology,  other fields besides the inflaton may be present during inflation and can have interesting consequences. 
Their dynamics can contribute to the inflationary mechanism 
 at the level of background or fluctuation evolution, and can leave imprints on  the properties of tensor modes,  for example by amplifying their spectrum. This implies that even models of inflation that normally predict a small value of $r$ can see the primordial tensor spectrum amplified by couplings with additional fields. 
 
 Perhaps the most studied example of multifields in inflation is the case with several scalar fields giving rise to a multiscalar  inflationary scenario (see e.g.~\cite{Gong} for a review and references therein).  However, spin one particles have also been considered in various scenarios (see e.g.~\cite{MSJS} for a review and references therein). 
Non-Abelian gauge fields have attracted a lot of attention in cosmology recently \cite{MSJ1,MSJ2,SJ,Adshead:2012kp,Adshead:2012qe,MNSJ}. Interestingly, in these models  the spin-2  sector  of the gauge field fluctuations  can provide a source for  primordial gravitational waves\footnote{Amplification  of  tensor  modes  by  spectator scalar fields has been discussed in e.g.~\cite{Biagetti:2013kwa,Biagetti:2014asa,Fujita:2014oba}, and the prospects for direct detection  with future gravitational-wave experiments were recently discussed in  \cite{Bartolo:2016ami}.  Amplification due to spectator axions coupled to  abelian gauge  fields has been discussed in \cite{Namba:2015gja,Peloso:2016gqs,ogan}.},   enhancing the amplitude of gravitational wave spectrum in single-field models of inflation, up to values observable with future CMB polarisation experiments. Moreover, the PGW  spectrum turns out to be chiral, making it potentially distinguishable from a vanilla inflation scenario \cite{Adshead:2013qp,Adshead:2013nka}.   These scenarios  are consequently very interesting because of their potential
 to generate distinctive observables that can be probed by the next generation of CMB polarisation experiments. 

An  $SU(2)$ gauge field $A_\mu^A$ was used in the so-called  {\em chromo-natural inflation} (CNI) model proposed in \cite{Adshead:2012kp,Adshead:2012qe}. The original motivation of this model was to relax  the requirement of a super-Planckian decay constant $f$  in natural inflation \cite{NI}, by using the Chern-Simons  (CS)  coupling 
$\chi  F_{\mu\nu}^A\tilde F^{A\,\mu\nu} $ of the axionic inflaton field $\chi$, to the gauge field with field strength $F_{\mu\nu}^A$.   This coupling effectively adds a friction term to the axion dynamics, allowing inflation to occur in a steeper potential with $f<\Mp$. 
It has however been demonstrated   that in its original form the CNI model is not observationally viable \cite{Adshead:2013qp,Adshead:2013nka}. An interesting proposal to alleviate this problem is to introduce a separate inflationary sector, while the gauge-axion sector acts only as a {\em spectator} \cite{OS,DFF}:  we dub this scenario spectator CNI (SCNI). 
A common feature of both these proposals,  which shall constitute an important point in our investigation, is the need for a  large axion-gauge  CS coupling (where the axion, gauge field -- and inflaton in SCNI -- are canonically normalised): %
\beq\label{Lpheno}
{\mathcal L} \supset \frac{\lambda}{4 f} \chi F_{\mu\nu}^A\tilde F^{A\,\mu\nu} \,,
\eeq
where $F=dA - g_A A\wedge A$, $g_A$ being the gauge coupling. Successful chromonatural inflationary proposals  that enhance tensor modes  
can be constructed, but  have to address the following theoretical challenges:
\begin{enumerate}
\item 
The typical values for the couplings appearing in eq.~\eqref{Lpheno}  that ensure a consistent background 
evolution are\footnote{In \cite{Azadeh}, a possibility to get $\lambda/f \sim {\mathcal O}(10)/\Mp$  in CNI was presented. We will come back to this point later.}  
\beq\label{Lpheno2}
\lambda\,\frac{\Mp}{f} \gtrsim 10^{4}\,,
\eeq
 which can be achieved with a large CS coupling $\lambda$, and/or a smaller-than-Planck decay constant\footnote{E.g.~$(\lambda, f, g_A) =(200,0.01\Mp, 2 \times 10^{-6})$ in \cite{Adshead:2012kp},  $(\lambda, f, g_A) =(2000,0.2\Mp, 10^{-3})$ in \cite{Adshead:2013nka};  $(\lambda, f, g_A) =(500,0.01\Mp, 1.11 \times 10^{-2})$ in \cite{DFF}.} $f$. 
However, as noted  already in \cite{Reece17,Reece}, for canonically normalised axion and gauge fields, the CS coupling is related to the gauge coupling $g_A$ as $$\lambda = \frac{k g_A^2}{8\pi^2}\,,$$ where $k \in {\mathbb Z}$. Given that the gauge coupling $g_A$ needs to be smaller than one
 to ensure perturbative control, %
 a large CS coupling  requires $k \gg 1$, and/or $f\ll \Mp$, which are hard to achieve, as discussed in  \cite{Reece}. 
\item An enhancement of  primordial tensor modes  generally implies that the amplitude of scalar or vector fluctuations is much amplified. One needs to avoid excessive backreaction of  the   gauge field  fluctuations to the background -- as  emphasised for example in \cite{DFF,Fujita17,MK} --  that would spoil inflation.
This condition typically requires {\it very small values} for the gauge coupling $g_A$, which somehow is in contrast 
with the condition of large CS coupling $\lambda$ met in the previous point. 
\end{enumerate}

 These facts indicate that it is not straightforward to design theories with the correct  properties  to realise   satisfactory (S)CNI models, and one has to carefully balance between the different  requirements on the quantities involved.

In this paper we  explore the possibility of embedding SCNI in string theory, which may offer new parameters that  provide the required conditions for a consistent  SCNI model, accompanied by a relatively large amplitude of the gravitational wave spectrum and a controllable backreaction from the gauge tensor fluctuations.

\subsection*{Brief review of attempts so far to embed (S)CNI in supergravity and string theory}
Before discussing our  model, we  review two ideas that have been proposed so far in this direction.  

The first  interesting possibility is to embed CNI in supergravity, as proposed   in \cite{DAgata}. In supergravity, the coupling between the axion (and saxion) to the gauge sector is dictated by the {\em gauge kinetic function}, $f_A$, as:
\beq
\label{Lsugra}
{\mathcal L} \supset - \frac{{\rm Re}(f_A)}{4} F_{\mu\nu}^A F^{A\,\mu\nu} + \frac{{\rm Im}(f_A)}{4} F_{\mu\nu}^A\tilde F^{A\,\mu\nu} \,,
\eeq
where $f_A$ is a holomorphic function of the superfields, including the axion that acts as the inflaton in CNI. 
The proposal to account for a large CS coupling was to introduce two canonically normalised  (super)fields, $\Phi_j= \alpha_j +i \phi_j$, $j=1,2$ and choose the gauge kinetic function as\footnote{In \cite{DAgata}, $f_A= 1+  c\,\frac{\Phi_1}{\Phi_2}$ with the norm  $|c|$ a number of order one; $c$  needs  to be purely imaginary in order for the axion-gauge coupling to be non-zero along the inflationary trajectory, which is given by ${\rm Re} \,\Phi_i=\alpha_i=0$.}  $f_A= 1+ i c\,\frac{\Phi_1}{\Phi_2}$ with  $c$ a number of order one. The inflationary trajectory occurs for $\alpha_i=0$, such that,  Re$(f_A)=1$, while Im$(f_A)=\phi_1/\phi_2$. The superpotential is chosen such that $\phi_2$ is  heavier  than the Hubble scale and it is thus kept fixed at its minimum at a suitable value of $\langle\phi_2\rangle$ during inflation\footnote{The K\"ahler potential and superpotential for this model are given by $K=\frac12 (\Phi_1 + \bar\Phi_1)^2 + \frac12 (\Phi_2 + \bar\Phi_2)^2 + S \bar S$ and $W = S \left[i \sqrt{2}\Lambda \sinh{\left(i \Phi_1/\sqrt{2}\Phi_2  \right)} + (\Lambda_2^2 + \Phi_2^2)\right]$, where $\Phi_i= \alpha_i +i \phi_i$ and $\alpha_i=S=0$ during inflation, while $\Lambda_2=\langle\phi_2\rangle$ sets the large coupling required.}. As we have mentioned before, CNI has been shown to be observationally unviable, however, one could  in principle add to this set-up  an inflationary sector as in \cite{DFF}, while keeping the axion-gauge sector as spectators. However, the viability of such a model will need to be carefully scrutinised when adding more fields; also  the backreaction of the gauge field tensor fluctuations on the background evolution will be hard to  control without introducing new parameters (i.e.~superfields)\footnote{As we will discuss, the gauge coupling needs to be sufficiently small to guarantee control on the backreaction. This could be introduced in supergravity by adding a third superfield, $\Phi_3$ to the configuration of \cite{DAgata},  in addition to the inflationary sector, such that $\langle\Phi_3\rangle$ sets the required small coupling. } \cite{DFF,Fujita17,MK}. We will discuss this aspect in detail in this paper. 

\smallskip

The second possibility  is the work of  \cite{McDA} to 
  embed SCNI in a string theory set-up. In  \cite{McDA}  a simplified string model is considered using gaugino condensation on magnetised D7-branes in type IIB CY orientifold compactifications, and  the axion associated to the 2-form potential  $C_2$ present  in the compactification (this  was  used  in \cite{Long,Ido} to realise natural inflation in string theory). It is argued that this set-up could in principle be embedded in a large volume scenario (LV) for moduli stabilisation \cite{LV1,LV2}, with the inflationary sector being a model of K\"ahler inflation \cite{KI1}. Although an explicit   model was not  presented 
     they consider  as possible values for the relevant parameters:  
  $\lambda=50$, $f=10^{-12}\Mp$,  a gauge coupling   $g_A=0.7$, and an axion scale\footnote{Assuming an effective potential for the axion of the form $V=\mu^4(1-\cos{(\chi/ f)})$.}, $\mu\sim 5.89\times 10^{-8}\Mp$, that is, $\mu\gg f$, which may be theoretically inconsistent due to unitarity constraints as discussed in  \cite{Reece}. Importantly, the backreaction of the gauge field tensor fluctuations on the background was not considered in \cite{McDA}. It has been argued in the literature that  backreaction imposes strong constraints on the parameters \cite{DFF,Fujita17,MK}.  
  As we shall discuss in detail, keeping  under  control  the  backreaction of gauge fluctuations  requires an additional  parameter, which needs to be sufficiently small.

\subsection*{What we do in this work}

In  light of these results,
in this paper we consider in detail the requirements for an explicit realisation  of the SCNI scenario in a string theory set-up,  which can achieve the following three goals: 
\begin{itemize}
\item[1)]
 a successful background evolution, 
 \item[2)] a sufficiently large enhancement of the tensor fluctuations  which become chiral and potentially detectable   by future experiments,
 \item[3)]   a controllable backreaction from the tensor gauge fluctuations.
\end{itemize}
We  will show that  while 1) and 2) can be achieved using two available parameters in the set-up we consider,  3) needs a third parameter, present in our set-up. 

As pointed out in \cite{McDA} (and \cite{DAgata} in supergravity), string theory naturally includes the field content of SCNI, albeit in a more intricate form. Given the phenomenologically appealing features of this scenario, and the forthcoming experimental opportunities associated with CMB polarisation, it is of  paramount importance to investigate the possibility of realising this model in a theoretically consistent theory, while also keeping the backreaction under control. 
 
\smallskip

As a benchmark  set-up, we adopt the LV moduli stabilisation scenario \cite{LV1,LV2} in  type IIB CY orientifold compactifications. Within  this framework, we consider   K\"ahler inflation, where the tensor-to-scalar ratio is too  small to be observationally relevant, namely $r\lesssim 10^{-7}$ \cite{KI1,KI2}. The minimal K\"ahler inflation model requires three K\"ahler moduli: one of them acts as an overall volume, one acts as a stabiliser and the third does the job of the inflaton. In order to realise a SCNI scenario, we need to introduce a spectator sector to K\"ahler inflation. This requires a fourth K\"ahler modulus and gaugino condensation  on a multiply-wrapped magnetised D7-brane stack,  whose gauge field fluctuations couple to  a $C_2$  axion.  
This set-up  thus resembles the one  introduced in \cite{McDA}, however we will look at the full moduli stabilisation and cosmological evolution of the inflaton as well as the spectator sector, and thus explicitly identify the three necessary parameters (and order of magnitude) to realise the three goals stated above. 
Specifically, these parameters are:  the magnetic flux on the D7-brane stack,  the degree of the condensing group and the  wrapping number of this brane. 

\smallskip 

Working with  these tools, we organise our discussion as follows:
\begin{itemize}
\item
 In  section \ref{Sec2}, we discuss the dynamics of $SU(N)$ gauge fields in general multifield inflationary scenarios that  arise in typical supergravity and string theory models. In particular,  the  metric in the scalar manifold is generically curved, the scalar potential is non-separable, and the coupling of the scalars to the gauge fields is non-trivial. 
Within this  set-up, we discuss the background evolution and requirements for successful inflation in this more general configuration. Once specialised to the case of spectator chromonatural inflation, we show how the background evolution leads to a condition similar to  eq.~\eqref{Lpheno2}. 

\item
 In section \ref{Sec3} we introduce  K\"ahler inflation  and the specific set-up we use to realise SCNI in this framework, introducing the relevant parameters discussed above and presenting the full background evolution. 
  We use the results in section \ref{Sec2} to show that goals 1) and 2) discussed above can be realised using two specific parameters present in this configuration, namely,   magnetic flux on the D7 spectator brane, and the degree of the condensing group. 

\item %
In section \ref{Sec4} we focus on a  detailed analysis of  the tensor and scalar perturbations for the model of section \ref{Sec3}, including a careful estimate of the
amount of  backreaction induced by the gauge fluctuations. We show
that the amplitude of the primordial tensor modes can be enhanced by a factor of order 
 $10^{3}$ with respect to the typical values met in the standard K\"ahler inflation model:
 this factor
  is sufficient to make the spectrum detectable by future experiments. 
  In order to avoid excessive backreaction
from the gauge fluctuations, we show that a third parameter present in the model is needed, specifically, a non-trivial wrapping of the spectator D7-brane stack. 

\item  In section \ref{disc} we summarise and discuss  our findings, the viability of the parameters' values to realise the three goals above and  possible future directions.
\end{itemize}

\section{Gauge fields in general multifield inflation}\label{Sec2}

We start  by  introducing  the general set-up that arises when considering $SU(N)$ gauge fields coupled to an axion in a general multi(scalar)field inflationary system, and 
then  study in detail the requirements for successful inflationary dynamics.  Our aim is to
determine general conditions that a successful model of slow-roll inflation has to satisfy 
in the general set-up we study. Moreover, when specialising to the case 
spectator chromonatural inflation, we  show  that our formulas 
 require large Chern-Simons couplings between the axion and gauge fields 
 (as discussed around eqs.~\eqref{Lpheno} and \eqref{Lpheno2}).

\bigskip

The action describing the system we are interested in consists of multiple scalar fields interacting with a gauge field.
 It reads\footnote{In \cite{OS}, a generalisation of  CNI was presented where  the canonically normalised inflaton -- driven by a dilaton --   and the axion, are both coupled to a canonically normalised $SU(2)$ gauge field. In our notation, they had a flat scalar metric,  $\gamma_{ab} =\delta_{ab}$, $f(\phi^a)$  a function of the inflaton and $h(\phi^a) = \frac{\lambda}{f} \sigma$, where $\sigma$ denotes the axion.}
\beq\label{action2}
\!\!S=\! \int{\d^4x \sqrt{-g} \left[\frac{\Mp^2}{2} \,R  - \frac{\gamma_{a b}(\phi^c)}{2} \partial_\mu \phi^a\partial^\mu \phi^b -V(\phi^a)  - \frac{f(\phi^a)}{4} F^A_{\mu\nu}F^{A\,\mu\nu} + \frac{h(\phi^a)}{4} F^A_{\mu\nu}\tilde F^{A\,\mu\nu}  \right]}\,,
\eeq
where  $\gamma_{ab}(\phi^c)$ is the metric of the scalar manifold spanned by the scalar fields $\phi^a$, and $V(\phi^c)$ is the scalar potential, which is generally  not separable; that is, the scalar fields generically interact via the kinetic and/or potential terms. The scalar sector with $a=1,\dots n$ scalars, contains the axion as well as the inflaton(s), which may be constituted by one or more fields, non-trivially coupled to one another.  
The coupling of the scalars (inflaton(s) and axion(s)) to the gauge  sector  is dictated by the  functions $f(\phi^a), h(\phi^a)$, which  generically depend on the scalar fields\footnote{As can be inferred from the introduction, and as  we will discuss below, they are related to the real and imaginary parts of the {\em gauge kinetic function}.}.

The gauge group is in general $SU(N)$ and the gauge field strength,  $F^A_{\mu\nu}$, is given by 
\beq
F^A_{\mu\nu}=\partial_{\mu}A^A_{\nu}-\partial_{\nu}A^A_{\mu}-  f^{A}_{BC}A^B_{\mu}A^C_{\nu} \,,
\eeq
where $f^{ABC}$ are the structure functions of $SU(N)$, and 
the dual, $\tilde F^{A\,\mu\nu}$,  is defined as $\tilde F^{A\,\mu\nu} =\epsilon^{\mu\nu\alpha\beta}F^A_{\,\alpha\beta}/(2\sqrt{-g})$ with $g$ the metric determinant. 
Let us  stress  that at this stage the gauge field is not canonically normalised. Thus,  there is no gauge coupling,  $g_A$, appearing in the definition of $F^A$, nor in the action. The gauge coupling is field-dependent and it is given by $g_A^2 = 1/f(\phi^a)$ once  the scalar field $\phi^a$, coupled  to the gauge field, has been stabilised.  However, while the scalar is evolving, we can define an {\em instantaneous gauge coupling} at a fixed time, $t_0$ as,  $g^2_{A,0} = 1/f(\phi^a (t_0))$ as will be the case in our string theory set-up. 

We now discuss the cosmological background evolution and slow-roll dynamics of the system. Specifically we are interested in the case where  two scalar fields couple to the gauge field via the functions  $f(\phi^a)$ and $h(\phi^a)$, plus an additional scalar field, not coupled to the gauge field, which  acts as the inflaton. That is, the spectator sector consists of two scalars coupled to the gauge field. This system generalises the  chromonatural inflation models discussed in \cite{Adshead:2012kp,DFF}.

\subsection{Background evolution and slow-roll inflation}

We consider a homogeneous and isotropic flat FRW metric $ds^2 = -dt^2 + a(t)^2dx^idx_i$, where $a(t)$ is the scale factor.
To treat the $SU(N)$ case, we proceed as follows. First,  homogeneity and isotropy of the gauge field energy density, and hence the background can be maintained by  splitting the $SU(N)$ gauge group into $\mathcal{N} = [N/2]$ ($N/2$ mod 2) disjoint sub-groups of $SU(2)$ \cite{Caldwell:2017chz, Caldwell:2018feo}. In each sub-group, the gauge field equals a different scalar, which can be  assumed to be locked into the following isotropic configuration,
\beq\label{gfA}
A^{A}_{0(n)} = 0,~~~~~~A^A_{i(n)} = a(t)Q_n(t)\delta^{A}_i,~~~~~~~~~~ n=1,\dots,\mathcal{N}\,.
\eeq
Notice that now within each sub-group, $f^{ABC} = \epsilon^{ABC}$. In the background configuration  \eqref{gfA}, the field strength tensor components within each sub-group are,
\bea
F^{A}_{0i(n)} &=& -a E^A_{i(n)} = a(t)\left[H Q_{(n)}(t) + \dot{Q}_{(n)}(t)\right]\delta^{A}_i,\\
F^{A}_{ij(n)} &=& a^2 \epsilon_{ijk} B^{A}_{k(n)}=- \,\epsilon^{A}_{ij} ~\big[a(t) Q_{(n)}(t)\big]^2\,,
\eea
where $H=\dot a/a$ is the Hubble parameter. 

For $\mathcal{N}>1$, it is difficult to solve the full background evolution.  However, a simplification occurs if one assumes that  each $SU(2)$  sub-group has a common field strength\footnote{This can be realised by initialising the system with a common initial condition. See also the discussion in \cite{Caldwell:2017chz}.}, $F^A$.  In this case, the system and equations of motion are equivalent to those of the single $SU(2)$ case by replacing $A_i^A\to A_i^A/\sqrt{\cal N}$, that is, 
\beq
Q\to \fr{Q}{\sqrt{\mathcal{N}}}\,, 
\eeq
and defining   an {\em effective gauge coupling}, $ {\tt g}$, as
\beq\label{geq}
{\tt g} \equiv \frac{1}{\sqrt{\mathcal{N}} } \,,
\eeq
which is now equivalent to having  $F=dA- {\tt g} \,A\wedge A$, without introducing this coupling into the function $h(\phi)$. Let us stress again that this is not the standard gauge coupling, which is still given by $g_A^2=1/f(\phi^a)$ defined above and at this stage is field dependent.  
From \eqref{geq}   we already see that a small (effective) gauge coupling $ {\tt g}$ can be achieved for a large gauge group, $N$: this fact will be important in what follows. 
Keeping this in mind, we focus on the single $SU(2)$ case with an effective gauge coupling given by \eqref{geq}, to investigate the background evolution of the system.

In this case, the  equations of motion for the metric are given by
\bea\label{Fried1}
3 \Mp^2 H^2 &=& \rho_\phi+\rho_{YM}, \\
2 \Mp^2 \dot H &=& -\dot\varphi^2 -  2f(\phi^a) \left[ \left(H Q+\dot{Q}\right)^2 + {\tt g}^2Q^4\right]  \label{Fried2} \, ,
\eea
where in \eqref{Fried1} the energy densities are  defined as
\beq
\rho_\phi =\frac{1}{2}\dot \varphi^2 + V(\phi_a) \,,\qquad  \rho_{YM} = \frac{3}{2}f\left[(HQ+\dot Q)^2  + {\tt g}^2 Q^4  \right]\,,
\eeq
and we have defined:
\beq\label{dvarphi}
\dot\varphi^2 \equiv \gamma_{ab}\dot \phi^a\dot\phi^b \,.
\eeq

\noindent The scalar  equations of motion are given by  
\beq\label{scalar1}
\ddot{\phi}^a+3 H \dot{\phi}^a+\Gamma^a_{bc}\,\dot{\phi}^b\dot{\phi}^c+\gamma^{ab}\,V_{,b}=-3 {\tt g}\, \gamma^{ab}h_{,b\,}Q^2\left(HQ+\dot{Q}\right)+\frac32 \gamma^{ab}f_{,b}\left(\left(HQ+\dot{Q}\right)^2-{\tt g}^2\,Q^4\right) \,,
\eeq
where   ${X}_{,a}$ denotes a derivative w.r.t.~to the field $\phi^a$. Note that the RHS of this equation will be zero for any scalar field that is not coupled to the gauge field. 
Note as well that these equations are valid also if the inflationary sector (decoupled from the gauge field) includes several scalars. In this case, there may be further interesting phenomenology due to possible large turns (see e.g.~\cite{Achucarro:2015rfa,Brown:2017osf,Garcia-Saenz:2018ifx,Bjorkmo:2019aev,Bjorkmo:2019fls,Aragam:2019omo,FatI}). We leave exploration of this possibility for future work. 

Finally, the gauge field equation is given by 
\bea\label{GF1}
\ddot{Q}+  3H   \dot{Q}  + Q\left(\dot{H}+2 H^2 \right)
+2{\tt g}^2Q^3={\tt g}\, Q^2\, \dot{\phi}^a\frac{h_{,a}}{f}
-\dot{\phi}^a\frac{f_{,a}}{f}  \left(  QH +  \dot{Q}  \right)\, ,
\eea
where   on the RHS only the scalar fields coupled to the gauge field will appear. 

The energy-momentum conservation equation, $\nabla_\mu T^{\mu\nu} =0$,  where $T^{\mu\nu}$ is the total energy momentum tensor including the scalars and gauge field,  further gives ($\nu=0$):
\bea\label{econs}
&&\dot\rho_\phi +3H(\rho_\phi+p_\phi) = \dot \phi^a Q_{a\,YM} \,, \nonumber\\
&&\dot\rho_{YM} +4H \rho_{YM} = -\dot \phi^a Q_{a\,YM} \,,
\eea
with
\beq
Q_{a\,YM} \equiv -3\,{\tt g} \,h_{,a} Q^2(HQ+\dot Q) +\frac{3}{2}f_{,a} \left[(HQ+\dot Q)^2-{\tt g}^2Q^4\right]\,.
\eeq
Equations \eqref{econs} show in a neat way the non-trivial interplay between the scalar and  gauge field dynamics  in a cosmological setting.

\subsection{Slow-roll Dynamics}

To study inflation, we define the first slow-roll parameter in the usual way as
\beq\label{epsilon}
\epsilon\equiv  -\frac{\dot H}{H^2} = \epsilon_\varphi + \epsilon_E+\epsilon_B \,,
\eeq 
where we introduced the slow-roll parameters for the scalars and the gauge field:  
\beq\label{epsilons}
\epsilon_\varphi  \equiv \frac{\dot\varphi^2}{2H^2\Mp^2}  \,, \qquad
\epsilon_E  \equiv \frac{f(HQ+\dot Q)^2}{H^2\Mp^2}  \,, \qquad 
\epsilon_B  \equiv \frac{{\tt g}^2 Q^4 f}{H^2\Mp^2}  \,.
\eeq
During inflation,  $\epsilon= \epsilon_\varphi + \epsilon_E+\epsilon_B \ll 1$, which implies that $3\Mp^2 H^2\sim V$. Note that there may be different hierarchies among the individual slow-roll parameters, but the overall $\epsilon$ has to be small,  $\epsilon\ll1$. 

We  now define a second small  slow-roll parameter  as follows:
\beq\label{eta}
\eta \equiv \frac{\dot\epsilon}{H\epsilon}=
2\epsilon_H -2 \frac{\epsilon_\varphi}{\epsilon_H} \delta_\varphi + \xi_f (\epsilon_B+\epsilon_E) + 2\frac{\epsilon_E}{\epsilon_H} \delta_E + 4 \frac{\epsilon_B}{\epsilon_H} \delta_B\,\ll\,1\,,
\eeq
where we introduced  the quantities
\bea\label{deltas}
\delta_\varphi\equiv - \frac{\ddot\varphi}{H\dot\varphi} \,, \qquad 
\delta_E \equiv  \frac{(\dot Q+HQ)^{\dot{}}}{H(\dot Q+HQ)} \,,\qquad \delta_B\equiv \frac{\dot Q}{HQ} \,,
\eea
which are small during inflation (to ensure that $\eta$ is small barring cancellations), as well as the small parameter 
\beq\label{xif}
\xi_f \equiv \Mp \frac{ f_{,a}\dot\phi^a}{2f\dot\varphi} \sqrt{2\epsilon_\varphi}=\frac{f_{,a}\dot\phi^a}{2fH}\,.
\eeq
Notice  that the slow-roll conditions defined above do not involve the coupling between the axion and the gauge field, $h$. Therefore, this coupling  can be large during inflation.

Dropping all terms with time derivatives except on the RHS of the equation for $Q$, \eqref{GF1}, which transfers part of the scalar sector kinetic energy to the background gauge field,  we get the relation 
\bea
2H^2Q + 2{\tt g}^2Q^3& \simeq &{\tt g} Q^2 \frac{h_{,a}\dot \phi^a}{f} - HQ \frac{f_a\dot\phi^a}{f} \,. \label{Qsimp}
\eea
Introducing the key parameters
\bea\label{xihQ}
\xi_h \equiv \frac{\Mp}{f}\frac{ h_{,a}\dot\phi^a}{2\dot\varphi} \sqrt{2\epsilon_\varphi} = \frac{h_{,a}\dot\phi^a}{2f H}\,, \qquad \qquad\quad 
\xi_Q  \equiv \frac{{\tt g}\,Q}{H} \,, 
\eea
eq.~\eqref{Qsimp} becomes 
\bea
1+\xi_Q^2 \simeq \xi_h\, \xi_Q - \xi_f  \label{xiseq}   \,.
\eea
 
\bigskip
\noindent As we saw, the slow-roll conditions imply that $\xi_f \ll1$, while no condition is required for $\xi_Q, \xi_h$, which can be significant, and given \eqref{xiseq}, we have, $ \xi_h \gtrsim \xi_Q$. 
Stability analysis in CNI have shown that scalar perturbations are stable for $\xi_Q>\sqrt{2}$ and therefore $\xi_h\gtrsim\xi_Q>\sqrt{2}$ \cite{Adshead:2012kp,Adshead:2012qe,DP,Azadeh}.
Then  $\xi_h > 1$, which implies 
\beq\label{xiconstr}
\frac{\Mp}{f}\frac{ h_{,a}\dot\phi^a}{2\dot\varphi} > \frac{1}{ \sqrt{2\epsilon_\varphi}}.
\eeq

\bigskip
\noindent 
This is the first important constraint on the background field evolution that any successful model of inflation
described by action \eqref{action2} should satisfy. 
Note that this relation depends on the ratio between the gauge field coupling to the scalars via $h_{,a}/f$  and has non-trivial implications for their values as we now discuss.  

\subsection{Large Chern-Simons couplings are needed in spectator chromo-natural inflation}

Let us consider  the condition in eq.~\eqref{xiconstr} in the known cases of CNI and SCNI. In those examples, $h(\phi^a) = \lambda \chi/f $ (see eq.~\eqref{Lpheno} in the Introduction) where $\chi$ is the axion, while $f(\phi^a)=1$. Then,  condition \eqref{xiconstr} reduces to \cite{Azadeh}
\beq\label{xiSCNI}
\frac{\Mp}{f}\frac{ \lambda}{\sqrt{2} }>\frac{1}{\sqrt{\epsilon_\chi}}\,,
\eeq
where $\epsilon_\chi$ is the slow-roll parameter associated to the axion field alone: $\epsilon_\chi = \frac{\dot\chi^2}{2\Mp^2 H^2}$ (that is, the inflaton does not enter in this relation in the case of SCNI). Therefore it is clear that in CNI, where the axion is also the inflaton, the larger the combination $\lambda\,\Mp/f$ is, the smaller $\epsilon_\chi$  will be. In other words, to avoid an excessively large coupling $\lambda\,\Mp/f$, $\epsilon_\chi$ will be as large as possible (see e.~g.~\cite{Azadeh} where $\Mp\lambda/f\sim 10$). On the other hand, in order to ensure  a maximal  enhancement of the  tensor fluctuations from the gauge field, the parameter $\xi_Q$ needs to be as large as possible, while keeping the backreaction of the gauge tensor perturbations on the background equations of motion under control \cite{DFF,Fujita17}. We will see this in detail in the string theory realisation in sections \ref{Sec3} and \ref{Sec4}.

In the extended version of chromonatural inflation  \cite{DFF}, the axion and gauge fields act as  spectators and therefore the requirement of a large axion coupling from the condition \eqref{xiSCNI} can be relaxed by consistently maximising the value of $\epsilon_\chi$, while   having  $\epsilon_\phi< \epsilon_\chi\ll 1$, where $\phi$ is the inflaton.  An extreme example of this situation with $\epsilon_\phi\sim 10^{-73}$, $H_{infl}\sim10^{-40}\Mp$ and $\epsilon_\chi\sim10^{-6}$ (with  $\lambda/f\sim 10^{4}\Mp^{-1}$, $\xi_Q\sim 44$, $g\sim 10^{-36}$)
 was discussed in \cite{Fujita17}\footnote{See however \cite{PPU} for restrictions on the model in \cite{Fujita17}.}.

In the  rich multifield  model we consider in the next section, there is not a  neat distinction between each scalar's dynamics, and therefore, $\xi_h$ is  dictated by the evolution of all scalar fields,  as seen in  \eqref{xihQ}. This means that this large coupling requirement could  be mildly relaxed   due to the multiple field evolution\footnote{As all the scalar fields interact non-trivially via the kinetic and potential terms, non-linearities may restrict further the parameter space along the lines discussed in \cite{PPU}. The details of this are outside the scope of the present paper and thus we will not consider this issue in this paper.}. Moreover, as we already mentioned, we are also aiming to maximise the tensor fluctuations' enhancement, that is largely determined by the magnitude of $\xi_Q$, as well as keeping the backreaction under control.

\section{Towards Spectator Chromonatural  K\"ahler  Inflation}\label{Sec3}

In this section we introduce the string theory set-up we consider  to embed a generalised SCNI model as described in the previous section. 
We start by introducing the original K\"ahler inflation set-up \cite{KI1} in type IIB Calabi-Yau orientifold compactifications within the large volume  moduli stabilisation \cite{LV1,LV2} framework. K\"ahler inflation  involves three K\"ahler moduli \cite{KI1} and  to this inflation sector we add the spectator sector given by  gaugino condensation on  multiply-wrapped magnetised D7-branes. The gauge fluctuations on these branes coupled to a  fourth  K\"ahler  modulus $T_4$ and a $C_2$ axion constitute  the field content of the spectator sector.  We also introduce a second stack of unmagnetised D7-branes wrapping same 4-cycle as the magnetised stack in order to allow for a suitable stabilisation of the fourth K\"ahler modulus, $T_4$. 
 This set-up is very close to that introduced in \cite{McDA} (see also \cite{Long,Ido} for similar set-ups used in the context of axion inflation in string theory). We will discuss briefly  the need to introduce a fourth modulus as well as a $C_2$ axion, rather than using the field content already present in the original K\"ahler inflation set-up. 

As we already mentioned in the introduction,  we are  interested in realising three goals with this model: 
1) a successful background evolution, 2) a sufficiently large enhancement of the tensor fluctuations in order to be observable by future experiments and 3) a controllable backreaction from the tensor gauge fluctuations. 
We will see that the construction described above includes three parameters that can be used to realise these goals, namely the brane magnetic flux, $m$, the gauge group degree\footnote{Not to be confused with the e-folds number in what follows!}, $N$ and the wrapping number, $n$. In this section we  consider suitable  values for  these parameters that  allow us to achieve the goals above 
and  use numerical methods to study   the background configuration and  cosmological predictions for the host inflationary model. In  the next section \ref{Sec4} we look the cosmological perturbations and backreaction for the model.
 
\subsection{Set-up and parameters} 
The four dimensional  effective action  arising  in models of inflation in string theory  such as K\"ahler Moduli Inflation is given by  

\beq\label{action1}
\!\!\!\!S \!\!=\!\! \int{\d^4x \sqrt{-g} \left[\frac{\Mp^2}{2} R  -   K_{i\bar j} \partial_\mu T^i\partial^\mu \overline{T}^{\bar j} -V(T^k)  - \frac{{\rm Re}(f_{A})}{4} F^A_{\mu\nu}F^{A\,\mu\nu} + \frac{{\rm Im}(f_{A})}{4} F^A_{\mu\nu}\tilde F^{A\,\mu\nu}  \right]},
\eeq
where $K_{i\bar j}$ is the metric of the scalar manifold spanned by the fields $T^i$, the Planck scale is given by $\Mp^2 = 4\pi {\cal V} M_s^2/g_s^2$, with ${\cal V}$ the  six-dimensional volume vev in string units, $M_s^2=((2\pi)^2\alpha')^{-1}$ the string scale and $g_s$ the string coupling. The $SU(N)$ gauge field(s) $F=dA - A\wedge A $,  give rise to gaugino condensation on  D7-branes, which generates a potential for the  K\"ahler moduli,   $T_i$. 
The other moduli present, namely the axiodilaton and  complex structure (and deformation moduli of D7-branes) are assumed to be already stabilised at the perturbative level by internal fluxes at a higher scale and integrated out consistently (see \cite{KI1} for more details).  
The coupling between the scalars (complex K\"ahler moduli) and the gauge field is given by the holomorphic {\em  gauge kinetic function}, $f_{A}(T^k)$.

Comparing \eqref{action1} to the general multifield action \eqref{action2} we introduced in section \ref{Sec2},  %
we see that we have a non-trivial scalar metric ($K_{i\bar j}$ here) mixing non-trivially the scalar fields; the scalar potential is in general a non-separable, non-trivial function of all the scalar fields, and there is a non-trivial well-defined coupling between the scalar fields and the gauge field.  
To make the comparison between \eqref{action1} in the form  \eqref{action2} more obvious, we can define  dimensionful fields as 
\beq\label{sfr}
\phi_a =\Mp  {\rm Re}\,T_i\,,
\eeq   
 and similarly for ${\rm Im} \,T_i$, such that the  kinetic term for the K\"ahler moduli is:
\beq
K_{i\bar j} \partial_\mu T^i\partial^\mu \overline{T}^{\bar j} = \fr{1}{2}\gamma_{ab} \partial_{\mu}\phi^a\partial^{\mu}\phi^b\,.
\eeq
We  discussed in the previous section how to deal with the general $SU(N)$ group in terms of an effective single $SU(2)$ group by introducing an effective gauge coupling \eqref{geq}, ${\tt g}=1/\sqrt{N/2}$. Let us stress again that this effective coupling is not the gauge coupling, which is field dependent and given by $ g^2_A=1/{\rm Re} {(f_A)}$ as  discussed in section \ref{Sec2}. 

\subsubsection{Scalar potential }

Let us start by discussing the scalar sector.
The scalar potential in \eqref{action1} is given by $\mathcal{N}=1$ supergravity in terms of 
 the K\"ahler potential and the superpotential as: 
\beq\label{Fterm}
V= \,e^{K/\Mp^2} \left(K^{i\bar j} D_iWD_{\bar j}\overline{W} - \fr{3}{\Mp^2}|W|^2\right)\,,
\eeq
where $D_iW=\partial_i W +K_i W$, $K_{i\bar{j}} \equiv \fr{\partial^2 K}{\partial T^{i}\partial \overline{T}^{\bar{j}}}$ and the K\"ahler  potential and  superpotential are given by  \cite{LV2}
\beq\label{KW}
\fr{K}{\Mp^2}= -2 \ln \left({\mathcal V} + \frac{\xi}{2 g_s^{3/2}} \right)-\ln(2/g_s)+ K_{\rm cs}\,,  \qquad  W =\fr{g_s^{3/2}\Mp^3}{\sqrt{4\pi}}\left( W_0 + \sum_i A_i e^{-a_i f_i} \right) \,,
\eeq
where $\mathcal{V}$ is the (Einstein frame) volume of the Calabi-Yau (CY) manifold $X$ measured in string units ($l_s = 2\pi \sqrt{\alpha'}$)  
and  $\xi = -\chi(X)\zeta(3)/(2(2\pi)^3)$ gives the  $\alpha'$ correction to the K\"ahler potential\footnote{The factor of $g_s^{3/2}$ arises because the volume is measured in Einstein frame \cite{LV1,LV2}.} \cite{CO,BBHL} where   $\chi(X)$ is the Euler characteristic number of the  CY manifold  and $\zeta(3)\simeq 1.2$. 
In eq.~\eqref{KW}, $K_{\rm cs}$ is the contribution to the K\"ahler potential from fixing the complex structure moduli to their minima and $W_0$  is the (tree-level) contribution to the superpotential coming from the fluxes after fixing the complex structure and axiodilaton to their minima. 
The sum over $i$ runs over the 4-cycles generating the non-perturbative contributions, where the $A_i$'s are model-dependent constants,
 while $a_i$ are given by  $a_i = 2\pi/N_i$ for gaugino condensation with gauge group $SU(N_i)$, and $f_i(T^k)$ are the holomorphic gauge kinetic functions.

The original LV scenario requires at least two K\"ahler moduli\footnote{It also requires more complex structure moduli than K\"ahler moduli $h^{(1,2)}>h^{(1,1)}$ where $\chi(X)=2h^{(1,1)} -h^{(1,2)}$ and hence $\xi>0$, in order to have the non-supersymmetric minimum at large volume, where the leading contribution to the scalar potential coming from the $\alpha'$ correction needs to be positive \cite{BaBe,LV1,LV2}.}: a `large' modulus controlling the overall volume and a `small' modulus corresponding to a blow-up cycle. K\"ahler moduli inflation within the LV framework requires  at least one more K\"ahler modulus, which can then act as the inflaton \cite{KI1}. The volume has the form: 
\beq\label{KIV}
{\mathcal V} = \alpha \left(\tau_1^{3/2}-\sum_{i=2}^n\lambda_i\tau_i^{3/2} \right)
\eeq
so that here $\tau_1$ is the large cycle, while $\tau_i$, $i=2,\dots, n$ are small blow-up cycles. The parameters $\alpha, \lambda_i$ are model-dependent constants that can be computed once a particular CY has been identified. 

In the minimal set-up of K\"ahler inflation with three K\"ahler moduli, $\tau_i$, the volume is given by \eqref{KIV} with $\tau_1\gg \tau_2,\tau_3$. A detailed  analysis of the scalar potential shows that it is possible to focus on a two-field evolution \cite{KI2}, where the inflaton $\tau_2$ and its axionic partner $b_2$ evolve. Therefore a natural question is whether their coupling to the D7-brane gauge field(s) can provide a source for primordial gravity waves, realising the SCNI  discussed in  \cite{DFF} in this minimal 3-field set-up. In this case the gauge kinetic functions are simply given by $f_i=T_i=\tau_i + i b_i$, so one could imagine an example with the real part of one of the K\"ahler moduli (and gauge kinetic function), say $T_2$, drives inflation, while its axionic part (and imaginary part of the gauge kinetic function) determines the coupling to $F^A\tilde F^A$. However, in this case it is clear that the required large coupling between the axion and the gauge field cannot be achieved (see eqs.~\eqref{xihQ}, \eqref{xiconstr}).
We therefore introduce a  suitable spectator sector, while  the real part of $T_2$ ($\tau_2$) drives inflation and its axionic partner ($b_2$) is stabilised as in the original model of K\"ahler moduli inflation \cite{KI1}.

\subsubsection{Spectator  sector}

The spectator sector  we introduce arises from a multiply-wrapped magnetised D7-brane stack along a 4-cycle parameterised by a fourth K\"ahler modulus $T_4$. 
In this case, the gauge kinetic function becomes \cite{Long,JL1,JL2,GKPW,McDA}
\beq\label{gkfFlux}
f_4 = n\left(T_4+ \kappa^A_{bc} \,G^b {\tt f}^c
+ \frac{ \kappa^A_{bc} \,{\tt f}^b{\tt f}^c}{2g_s}\right) \,, 
\eeq
where $n$ is the wrapping number, $\kappa^A_{bc}$ are  intersection numbers, ${\tt f}^c$ is the D7-brane magnetic flux and  $G^a$  is given by\footnote{Where we have assumed that the axiodilaton imaginary part has  been fixed to zero.}
 $G^a = \frac{1}{g_s} {\tt b}^a + i {\tt c}^a$, where ${\tt b}^a$ and ${\tt c}^a$ are the axions descending  from the $B_2$ and $C_2$ forms present in the theory. The K\"ahler coordinate $T_4$ is  shifted by $G$ as $T_4 \to T_4 -\frac{g_s}{4} \kappa^A_{bc} \, G^b(G+\bar G)^c$  \cite{JL1,JL2,GKPW}. The magnetisation of the D7-branes also contributes to the D-term for the D7-brane gauge theory \cite{JL2,GKPW}. In general this can receive contributions from matter fields living on the D7-branes. Here we  assume that these have been stabilised at a high scale together with the complex structure moduli and the axiodilaton. Furthermore, we assume that the D-terms also contribute to the stabilisation of the $B_2$ axion at  ${\tt b}=0$ \cite{Long,McDA}.

Finally, for a successful stabilisation of   $\tau_4$ consistent with our generalised SCNI scenario, we  introduce a second stack of unmagnetised D7-branes wrapping the same cycle as the spectator brane, which gives rise to a second non-perturbative contribution for the modulus $T_4$. That is, the superpotential  in \eqref{KW} includes two terms for the spectator sector given by 
\beq\label{W4}
W_{np}^{\small (s)} = A_{4} e^{- a_{4} T_{4}} +   A e^{- a  f_{4}},   
\eeq 
where 
\beq\label{gkfc}
f_{4}= n\left(T_4 + i \,m\, b \right)\,,
\eeq 
where  $T_4=\tau_4+ i b_4$ and we have denoted  the magnetic brane flux with $m$ and renamed the $C_2$ axion as $b$.

Let us summarise the configuration and parameters we have and compare them to the phenomenological DFF model \cite{DFF,Fujita17}. The spectator sector's action takes the following  form: 
\beq\label{LAsimp}
{\cal L} \supset -\frac{ f(\tau_4)}{4} F^A_{\mu\nu}F^{A\,\mu\nu} + \frac{ h(b)}{4} F^A_{\mu\nu}\tilde F^{A\,\mu\nu}\,,
\eeq
where\footnote{We have ignored here the shift in ${\rm Im} f_4$ due to $ b_4$, which will be stabilised during the cosmological evolution. We do this because the required value of the magnetic flux $m\gg \langle b_4\rangle $ and thus will not change the results. }
\beq\label{fh}
 f(\tau_4) = \tau_4, \qquad h(b) = m\, b, 
 \eeq
and we have absorbed the wrapping number $n$ into the gauge field $A^A$, whose field strength is now given by $F_2 = dA -{\tt g}\, A\wedge A$  with the effective  gauge coupling  redefined as
\beq\label{geq2}
{\tt g} = 1/\sqrt{n N/2}, 
\eeq
and as we have discussed, both $\tau_4$ and $b$ are dynamical during inflation. 
It is also important to remember that the scalar fields $\tau_4$ and $b$ are not canonically normalised. 
The scalar potential coming from \eqref{Fterm} for $b$ will take  the form (see below)
\beq\label{scalarb}
V(b) \sim  g(\tau_4) \cos\left(n\, a\, m\, b \right)\,,  
\eeq
where $g(\tau_4)$ is a function of the spectator saxion, $\tau_4$, and $a = 2\pi/N $. 

From our discussion in section \ref{Sec2}, we now see that the large coupling between the axion and the gauge field can be achieved by a large magnetic flux, which fixes $m$. We also need the gauge field to be sustained for a sufficiently long time in order to enhance the tensor spectrum, which requires a suitable value for the ``decay constant" of the axion. Note that since the scalar fields are not canonically normalised, the naive decay constant read off from the scalar potential \eqref{scalarb} 
$f_c = 1/(m \,n\,  a)$ is not correct. However we can identify an {\em instantaneous decay constant} as $f_c = \Mp\sqrt{\gamma_{bb}(\tau_4)}/(m\,n\, a)$  and as we will see, this needs to be of order $f_c\sim10^{-3}\Mp$, thus fixing the value of $m\,n\, a$. At this stage, we have two constraints and three parameters, $(n, m, N)$,  in order to have a successful inflationary evolution with a gauge field that can be sustained for long enough to enhance the tensor spectrum. 
However, besides a successful inflationary evolution, one needs to make sure that the backreaction of the tensor gauge perturbations are under control  \cite{DFF,Fujita17,MK,PPU}, which will require ${\tt g} \ll 1$, thus introducing  a third constraint and fixing the three parameters we have available.

 In summary, we have 3 parameters: $(m, N,n)$ to fix three constraints to fulfil  our goals 1), 2), 3)  discussed in the Introduction. In what follows, we will choose $m$, $a \,n \,m$  and ${\tt g}$ (thus fixing $(m,N,n)$)  to ensure a successful background evolution, large enhancement of tensor perturbations  and good control on the backreaction.

\subsection{Background evolution and cosmological parameters}

  The inflationary potential can be found from the superpotential and K\"ahler potential through \eqref{Fterm}, \eqref{W4}. The scalar manifold metric at leading order in $1/{\cal V}$ can be computed from \eqref{KW} to be 
    \beq
  K_{T_i\overline{ T}_{\bar j}} = \fr{3\alpha\lambda_i}{8\mathcal{V}\sqrt{\tau_i}} \Mp^2 \delta_{i\bar j}\,, 
  \eeq   
  while for the $C_2$ axion it has the form \cite{JL1,JL2}\footnote{The general form is given by $K_{a\bar b} = \frac{g_s}{{\cal V}}\kappa_{ab}^\alpha t_\alpha$ where $t_\alpha$ are 2-cycle volumes and are related to the 4-cycle volumes via $\tau^\alpha = \frac{1}{2}\kappa^{\alpha\beta\gamma}t_\beta t_\gamma$, while the volume can be written in terms of the $t_\alpha$'s as ${\cal V} = \frac{1}{3\!} \kappa^{\alpha\beta\gamma}t_\alpha t_\beta t_\gamma$. Here we assume that the there is an orthogonal basis such that $\tau_4\propto (t^4)^2$.}
  \beq
  K_{b\bar b} = \frac{g_s}{{\cal V}}\frac{\sqrt{\tau_4}}{\sqrt{\gamma}}\Mp^2\,,
  \eeq
 where $\gamma$ is a model dependent constant. The potential for $\tau_i$ and $b_i$ at large volume takes the form 
\beq\label{InfPot}
\hskip-0.5cm
V = \fr{e^{K_{\rm cs}}(g_s\Mp)^4}{8\pi}\left[ V_2 + V_4+ \fr{3\xi W_0}{4\mathcal{V}^3}+\fr{\beta}{\mathcal{V}^2}+V_3 \right]\,,
\eeq
where
\beq
V_2 = \frac{8 (a_2A_2)^2 e^{-2 a_2 \tau_2} \sqrt{\tau_2}}{3 \alpha\lambda_2 \mathcal{V}}+\frac{4 W_0a_2 A_2 e^{-a_2 \tau_2}\,\text{cos}\left( a_2 b_2\right) \tau_2}{\mathcal{V}^2}\,;
\eeq
\bea
&&V_4 =  \frac{8\tilde{a}^2 \tilde{A}^2\sqrt{\tau_4}}{3  \alpha \lambda_4 \V}e^{-\frac{2 \tilde{a}}{m}\tau_4} + \frac{16 \tilde{a} \tilde{A} a_4 A_4 \sqrt{\tau_4}}{3 \alpha \lambda_4 \V}e^{-\left(a_4 + \frac{\tilde{a}}{m}\right)\tau_4}\cos\left[a_4 b_4-\tilde{a}\left(b+\frac{b_4}{m}\right)\right] \nonumber \\
&&\hskip+1cm +\frac{4 \tilde{a} \tilde{A} W_0 \tau_4}{\V^2} e^{-\frac{\tilde{a}}{m}\tau_4}\cos\left[\tilde{a}\left(b+\frac{b_4}{m}\right)\right]+ \frac{8 (a_4A_4)^2 e^{-2 a_4 \tau_4} \sqrt{\tau_4}}{3 \alpha\lambda_4 \mathcal{V}}\nonumber \\
&&\hskip+1cm+\frac{4 W_0a_4 A_4 e^{-a_4 \tau_4}\,\text{cos}\left( a_4 b_4\right) \tau_4}{\mathcal{V}^2}
\eea
where we defined 
 \beq
 \tilde{a} \equiv a n m,  \qquad \tilde{A} \equiv \frac A m;
 \eeq
  the term proportional to $\beta$ in \eqref{InfPot} is an uplift term, taken to be of the form $V_{\rm uplift}=\beta/{\cal V}^2$ as in \cite{KI1}; and  $V_3$  comes from the stabilisation of the $T_3$ modulus and it is given by:
\beq
V_3 = \frac{8 (a_3A_3)^2 e^{-2 a_3 \langle\tau_3\rangle} \sqrt{\langle\tau_3\rangle}}{3 \alpha\lambda_3 \mathcal{V}}-\frac{4 W_0a_3 A_3 e^{-a_3 \langle\tau_3\rangle}\langle\tau_3\rangle}{\mathcal{V}^2}\,,
\eeq
with $b_3$ set to its minimum, $\langle b_3 \rangle = \pi/a_3$.  The small cycle $\tau_3$ acts as a stabiliser for the potential at large volume $\mathcal{V}$ so cannot be shifted far from its minimum. However  $\tau_2$, $\tau_4$ and $b_4$ can be displaced away from their minima leading to an effective three-field inflationary model. We have explicitly checked that the  $T_3$ modulus stays at its minimum with $\langle b_3 \rangle = \pi/a_3$ during the cosmological evolution in this case.  The $b_2$ axion can be started away from its minimum and its  effect on the inflationary evolution -- in particular, with $b_2$ well away from its minimum -- is to slow down $\tau_2$  further and inflation proceeds for longer. However, as shown in the original K\"ahler inflation scenario \cite{KI1},  $b_2$ can be set to its minimum, $b_2 = \langle b_2 \rangle = \frac{\pi}{a_2}$, where it is stable. 
  
  With this  system $\tau_4$ and $b$, coupled to the gauge sector, act fully as spectator fields. We require that both the kinetic and potential energy densities in the spectator fields are less than those of the inflaton. This ensures that the inflationary predictions of K\"ahler inflation driven by $\tau_2$ are not affected and thus $\tau_4$, $b$ and the gauge field, are true spectator fields.

\subsubsection*{Parameters}
We consider  the following set of parameters for the K\"ahler moduli\footnote{The parameters are the same as in example 4 in \cite{KI2}, with the addition of a fourth modulus and the $C_2$ axion. Moreover, as in \cite{KI1,KI2} we use $e^{K_{\rm cs}}$ to match the amplitude of the scalar power spectrum.}:
\bea\label{params2}
&& \xi=\frac12, \qquad \alpha =\frac{1}{9\sqrt{2}}, \qquad  \lambda_2=10, \qquad \lambda_3=1, \qquad \lambda_4=0.01, \nonumber \\
&& \nonumber \\
&& a_2=\frac{2 \pi}{30}, \qquad a_3=\frac{2 \pi}{3}, \qquad a_4= \frac{2 \pi}{50}, \qquad 
\taa =  40 a_4, \qquad g_s = 0.1, \nonumber \\
&& \nonumber \\
&& A_2=\frac{1}{1.7\times 10^6}, \qquad A_3= \frac{1}{425}\,, \qquad A_4=4.2 \times 10^{-9},  \qquad \tA = 0.0034A_4, \nonumber \\
&& \nonumber \\
&&   m= 10000, \qquad \gamma= 5, \qquad 
\beta = 6.94681\times10^{-5}, \qquad W_0=\frac{40}{17} \, .
\eea

Here $\lambda_4$ is chosen to be much smaller than $\lambda_2$ so that $\tau_4$ has lower kinetic energy, and therefore  a negligible contribution to $\epsilon_{\varphi}$. We require that the energy density of the universe receives only a small contribution from the spectator part $V_4$ so that inflation can proceed as expected with the important inflationary terms given in $V_2$. With this in mind, $A_4,\,a_4,\,\tA,\, \taa $ are chosen to be small relative to $A_2,\,a_2$. In addition to this, we would like the stabilisation of $\tau_4$ and $b_4$ to be dictated by the terms 
\beq\label{compV4}
\frac{8 (a_4A_4)^2 e^{-2 a_4 \tau_4} \sqrt{\tau_4}}{3 \alpha\lambda_4 \mathcal{V}}+\frac{4 W_0a_4 A_4 e^{-a_4 \tau_4}\,\text{cos}\left( a_4 b_4\right) \tau_4}{\mathcal{V}^2}\,,
\eeq
which is why $\tA$ is taken to be  small relative to $A_4$. Ensuring the terms involving $b$ are much smaller than the terms in \eqref{compV4} ensures that the $b_4$ axion will be minimised at $\langle b_4\rangle = \frac{\pi}{a_4}$, which in turn ensures that $\tau_4$ is stabilised at a much smaller value than $\tau_1$\footnote{Recall $\tau_4$ must be a `small' blow-up modulus, and a minimum at low values of $\tau_4$ is achieved through the terms in the potential of \eqref{compV4}. The minimum can be ruined if the terms involving the $b$ axion are too large and $\tau_4$ can be destabilised to very large values, $\tau_4 \gtrsim 10^3$.}. We are then left with some freedom in choosing $a$ and $m$, which are taken phenomenologically to lead to a successful enhancement of the PGW spectrum.

The global minimum of the potential for this set of parameters is found to be at:
\beq
\tau_1=2554.50, \qquad \tau_2=4.77523, \qquad \tau_3=2.65081, \qquad \tau_4=14.8743, \qquad  \mathcal{V} \rightarrow 10135.3 \, ,
\eeq
while the axions' minima lie at $b_i = \pi/a_i$. This is a relatively small-volume example for K\"ahler modulus inflation, however it is still consistent with the large volume approximation. With this value for the 6D volume,  the string scale results  $M_s\sim 3\times 10^{-4}\Mp$.

Displacement of $\tau_1$, $\tau_4$ and $b$ away from this minimum leads to a negligible shift in the values of $\tau_3$ and $\cal V$ at the new local minimum. Therefore, for numerical simplicity, we set $\tau_3$ and $\mathcal{V}$, as well as the $b_2, b_3$ and $b_4$\footnote{We explicitly checked that the $b_4$ axion is indeed stable at $\frac{\pi}{a_4}$ despite the extra terms in the potential that involve both $b$ and $b_4$. This is because these extra terms are chosen to be much smaller than the term proportional to $\cos(a_4b_4)$, which successfully stabilises the axion at $\frac{\pi}{a_4}$.} axions to their minima without loss of generality, so that we are  considering a three-field system $(\tau_2, \tau_4, b)$ with real-space field metric given by
\beq\label{gamma}
\gamma_{ab} = \begin{pmatrix}
	\frac{3 \alpha \lambda_2}{4 \sqrt{\tau_2} \V} &&0 && 0 \\
	0 && \frac{3 \alpha \lambda_4}{4 \sqrt{\tau_4} \V} && 0 \\
	0 && 0 && \frac{2 g_s \sqrt{\tau_4}}{\sqrt{\gamma} \V}
\end{pmatrix}.
\eeq 
Note that in the large volume limit, both the scalar manifold metric \eqref{gamma} and the scalar potential \eqref{InfPot} for the inflaton ($\tau_2$) and spectator sector ($\tau_4, b$) are decoupled. 

\smallskip
We now have all the ingredients of the low-energy action describing the dynamics of the inflaton and spectator sector fields. 
The equations of motion were discussed in Section \ref{Sec2}, and are given in equations   \eqref{Fried1}-\eqref{GF1}. 
The initial conditions for $\tau_4$, $b$, and $Q$, as well as the values of  the parameters are chosen phenomenologically to lead to a large (observable) enhancement of the gravitational wave spectrum (see section \ref{Sec4}) without leading to excessive backreaction from the gauge tensor perturbation (see section \ref{Sec6}). With this in mind, besides the parameters in \eqref{params2} we take $\g = \frac{1}{2000}$, which fixes the last parameter, namely, $n$. 

\begin{figure}[H]
	\centering
	\includegraphics[width=0.4\textwidth]{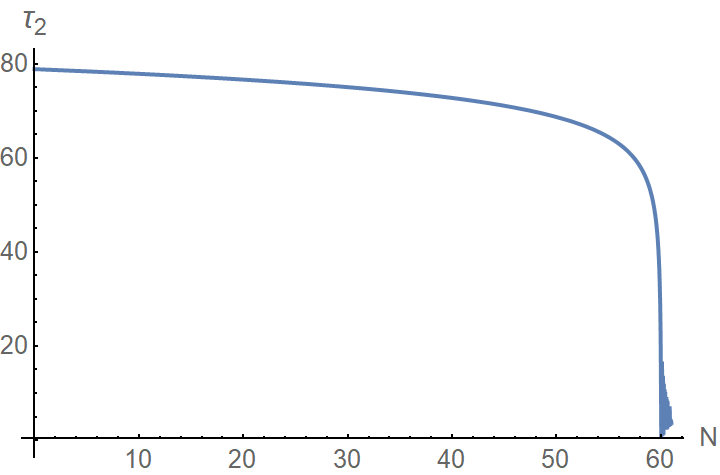}\qquad \includegraphics[width=0.4\textwidth]{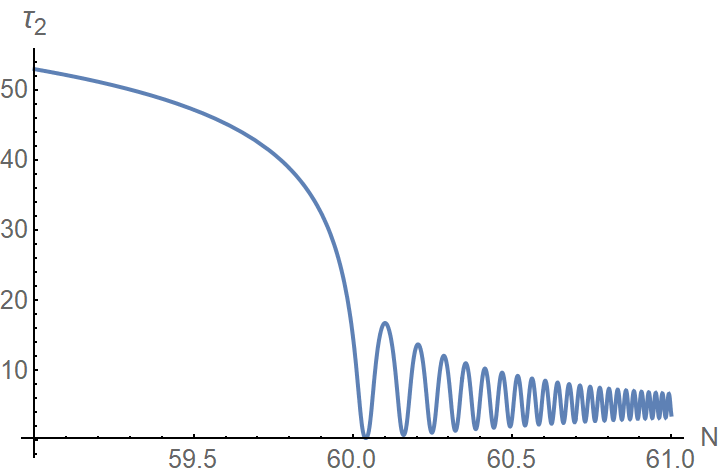}
	\caption{The evolution of the inflaton, $\tau_2$ during the last 60 e-folds of inflation (left) and during the last few e-folds (right). }
	\label{tau2KI2}
\end{figure}

\begin{figure}[H]
	\centering
	\includegraphics[width=0.4\textwidth]{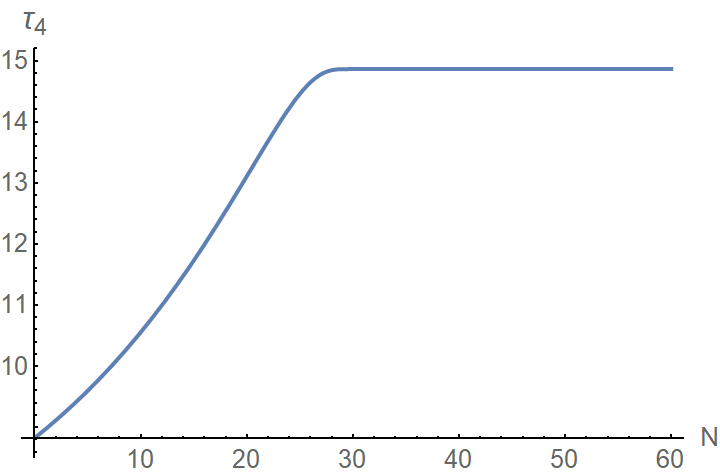}\qquad \includegraphics[width=0.4\textwidth]{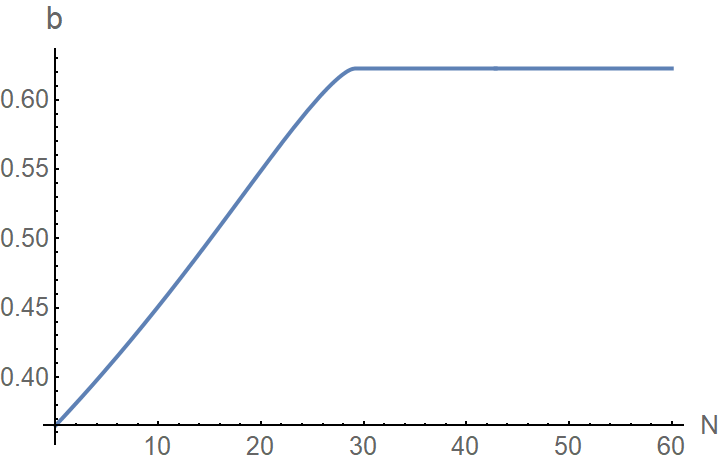}
	\caption{The evolution of the spectator modulus, $\tau_4$ and its axion partner $b$. Both fields reach their minima well before the end of inflation.}
	\label{tau4bKI2}
\end{figure}

\begin{figure}[H]
	\centering
	\includegraphics[width=0.4\textwidth]{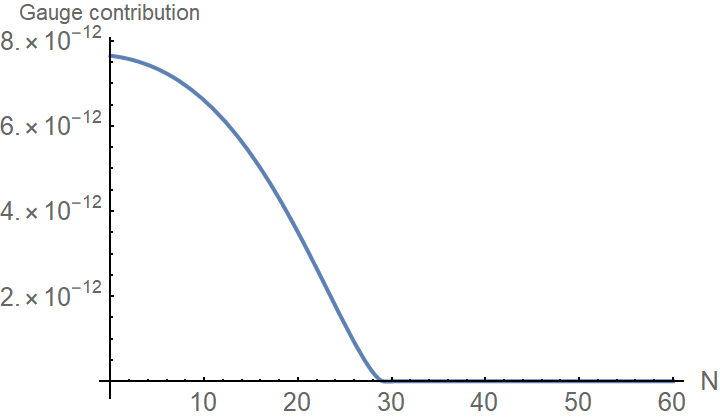}\qquad \includegraphics[width=0.4\textwidth]{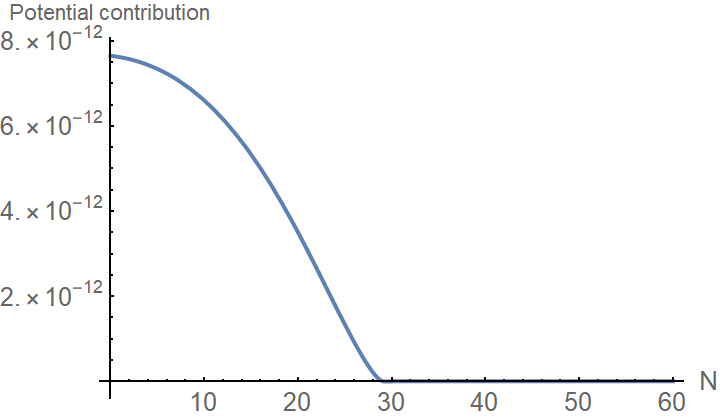}
	\caption{The contributions of (left) the term $-3\, \g\, \frac{m}{2\pi}\,\frac{\sqrt{\gamma} \V}{2 g_s \sqrt{\tau_4}}\,Q^2\,\left(HQ+\dot{Q}\right)$ provided by the gauge field, $Q$; and (right) the term $\frac{\sqrt{\gamma} \V}{2 g_s \sqrt{\tau_4}}\,\frac{\del V}{\del b}$ provided by the potential, to the equation of motion for $b$ given by the form in \eqref{scalar1}. The contribution from the gauge field almost exactly cancels the contribution from the potential leading to a slow-roll evolution for $b$.}
	\label{bcontrib}
\end{figure}

\begin{figure}[H]
	\centering
	\includegraphics[width=0.4\textwidth]{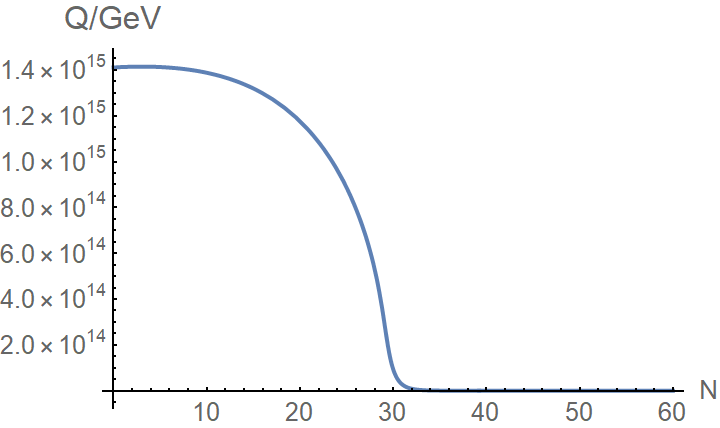}\qquad \includegraphics[width=0.4\textwidth]{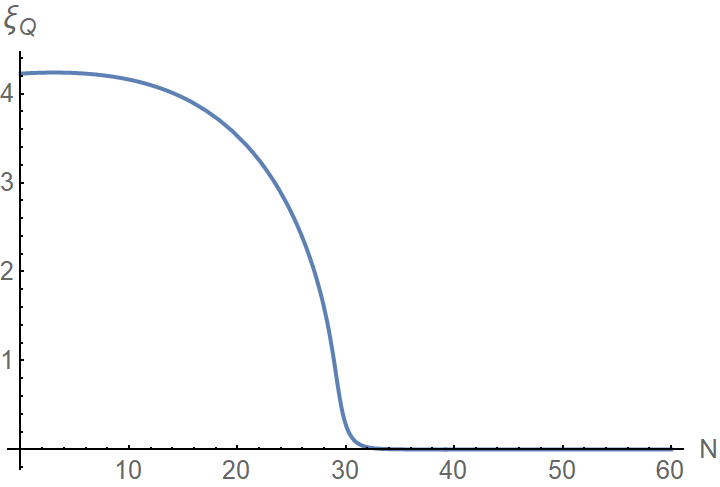}
	\caption{The evolution of the gauge field, $Q$ (left) and $\xi_Q = \frac{g Q}{H} $ (right) during the last 60 e-folds of inflation. The evolution of $Q$ is tied to the evolution of $b$.}
	\label{QKI2}
\end{figure}

\begin{figure}[H]
	\centering
	\includegraphics[width=0.7\textwidth]{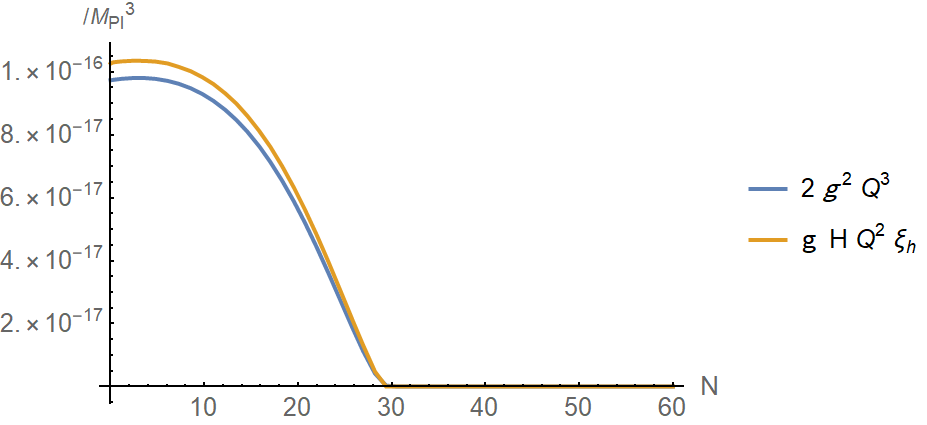}\qquad 
	\caption{The two dominant terms in the equation of motion for $Q$, \eqref{GF1}, satisfy $ 2 \g^2 Q^3 \sim \g H Q^2 \xi_h$. The term introduced through the coupling to the axion almost cancels $ 2 \g^2 Q^3$, a term that sends $Q$ to zero. In this way the axion-gauge coupling supports the gauge field. }
	\label{Qcontribs}
\end{figure}

\begin{figure}[H]
	\centering
	\includegraphics[width=0.4\textwidth]{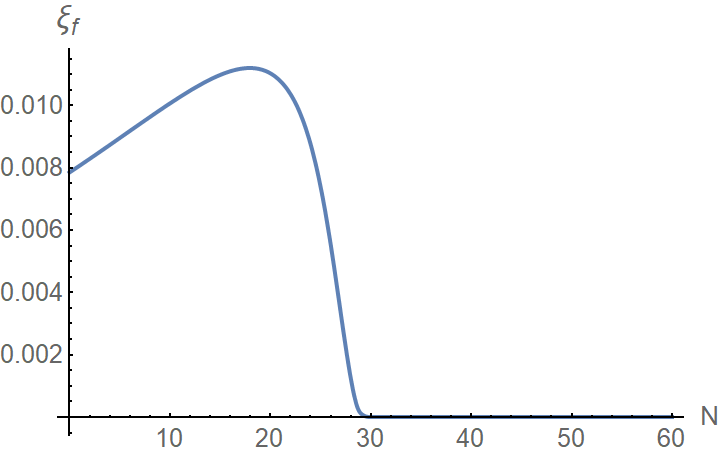}\qquad \includegraphics[width=0.4\textwidth]{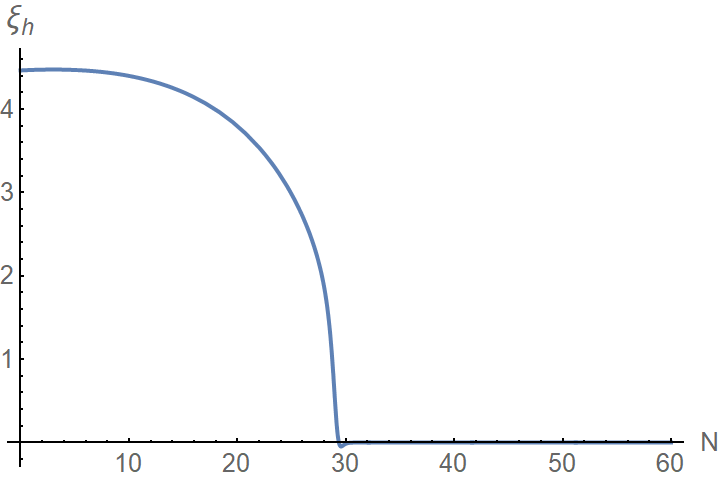}
	\caption{The evolution of the effective coupling constants of $\tau_4$ and $b$ to the gauge field, $\xi_f$ (left) and $\xi_h$ (right) during the last 60 e-folds.}
	\label{xifhKI2}
\end{figure}

\begin{figure}[H]
	\centering
	\includegraphics[width=0.4\textwidth]{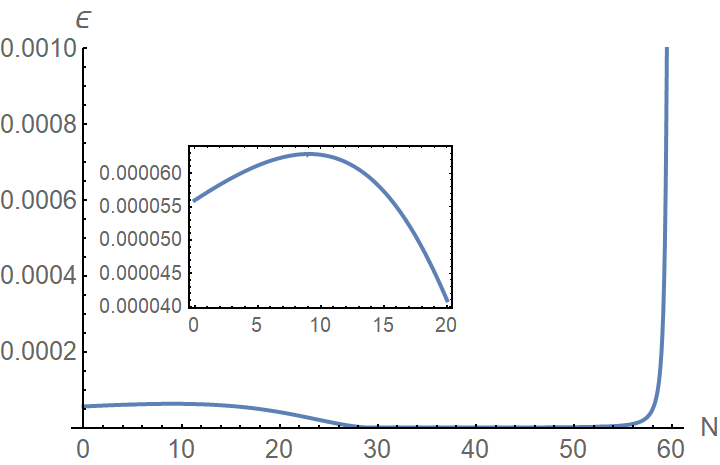}\qquad \includegraphics[width=0.4\textwidth]{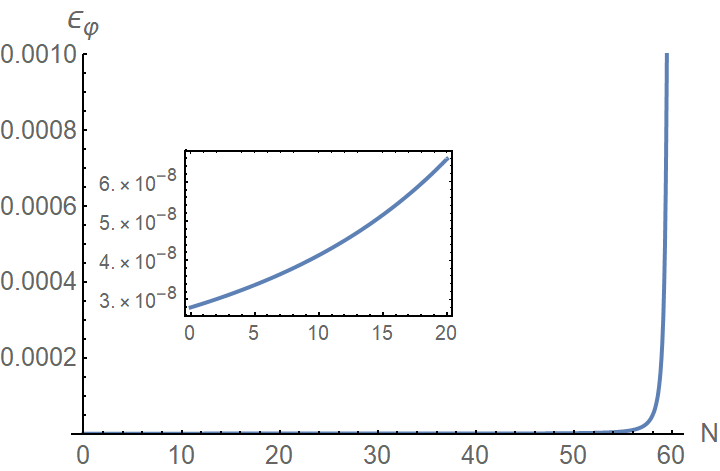}
	\caption{The evolution of the slow-roll parameter, $\epsilon$ (left), and the proportion of $\epsilon$ made up by $\epsilon_{\varphi}$, during the last 60 e-folds of inflation. }
	\label{epsKI2}
\end{figure}

\begin{figure}[H]
	\centering
	\includegraphics[width=0.4\textwidth]{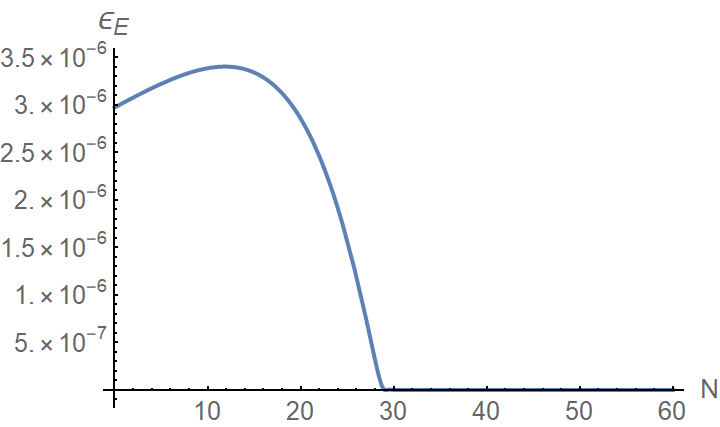}\qquad \includegraphics[width=0.4\textwidth]{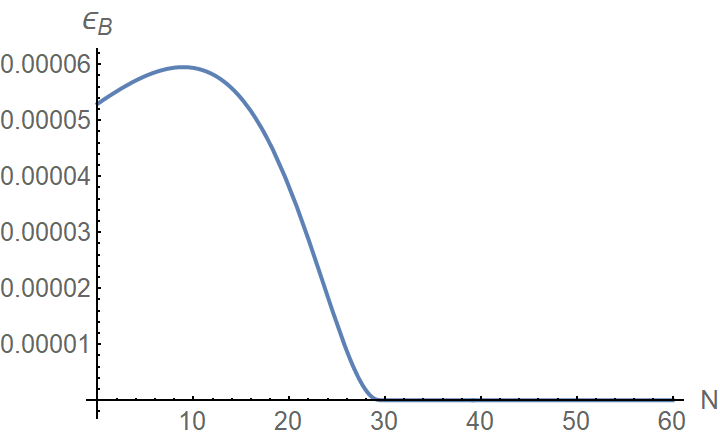}
	\caption{The evolution of the magnetic and electric components of the slow roll parameter, $\epsilon_E$ (left figure) and $\epsilon_B$ (right figure) respectively during the last 60 e-folds of inflation.}
	\label{epsgaugeKI2}
\end{figure}

\begin{figure}[H]
	\centering
	\includegraphics[width=0.4\textwidth]{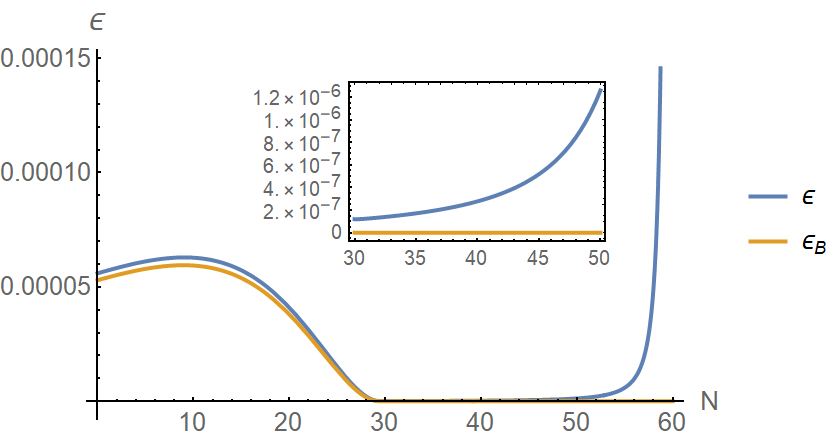}\qquad
	\includegraphics[width=0.4\textwidth]{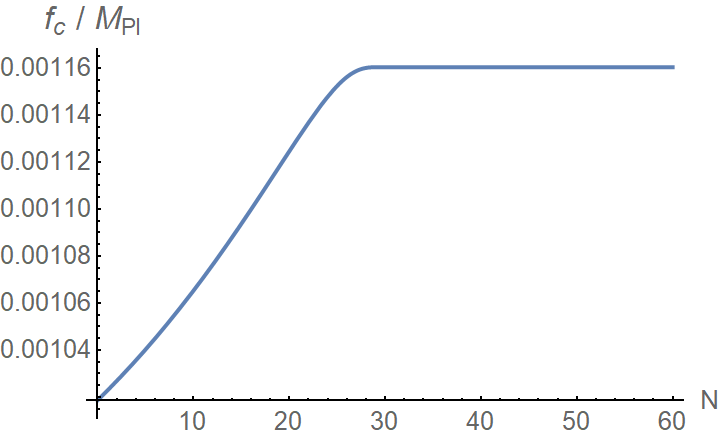}
	\caption{Left figure: Comparing $\epsilon_B$ with the overall slow-roll parameter, $\epsilon$. $\epsilon_B$ provides the largest contribution to $\epsilon$ for the majority of the last 60 e-folds of inflation. Right figure: Plot of the instantaneous decay constant, $f_c = \frac{\sqrt{\gamma_{bb}}}{a \,m\, n}$, during the last 60 e-folds of inflation.}
	\label{epsvBKI2}
\end{figure}

 The initial conditions are taken as:
\beq\label{int cons 2}
\tau_2=80.17, \qquad \tau_4=10, \qquad b=0.4 \fr{\pi}{a}, \qquad Q=8 \times 10^{-4} \Mp\, .
\eeq
The modulus $\tau_4$ is shifted slightly from its minimum while the axion, $b$, is well away from its minimum, but as can be seen in FIG. \ref{tau4bKI2}, as $b$ approaches its minimum $\langle b \rangle = \frac{\pi}{a}$, $\tau_4$ approaches its own $\langle \tau_4 \rangle =14.9$. The evolution of $\tau_4$ is thus quite trivial. The important spectator scalar field is of course the axion, $b$, which moves slowly towards its minimum as can be seen in FIG. \ref{tau4bKI2}. The axion's, $b$,  evolution is slowed considerably by its coupling to the gauge field, $Q$, which backreacts on it through the term on the RHS of \eqref{scalar1}, $-3\, \g\, \gamma^{ab}\,h_{,b}\,Q^2\left(HQ+\dot{Q}\right)=-3\, \g\, \frac{m}{2\pi}\,\frac{\sqrt{\gamma} \V}{2 g_s \sqrt{\tau_4}}\,Q^2\,\left(HQ+\dot{Q}\right)$. This term almost completely cancels with the potential term $\gamma^{ab}V_{,b}=\frac{\sqrt{\gamma} \V}{2 g_s \sqrt{\tau_4}}\,\frac{\del V}{\del b}$ as can be seen in FIG. \ref{bcontrib}. This ``slow-roll solution'' is the situation described and expected in chromonatural inflation \cite{Adshead:2012kp} and more generally in models with a spectator axion coupled to a gauge field \cite{McDA,DFF} where the gauge field, $Q$, has an attractor solution such that it forces the axion to roll slowly. This slow evolution of the axion  leads to the gauge field, $Q$, being sustained for a large period of time during inflation as is shown in FIG. \ref{QKI2}. The evolution for $Q$ satisfies $ 2 \g^2 Q^3 \sim \g H Q^2 \xi_h$ as can be seen in FIG. \ref{Qcontribs}. Left unchecked, the term $ 2 \g^2 Q^3$ will quickly send $Q$ to zero. The term in eq.~\eqref{GF1} due to the coupling between the axion and gauge field almost cancels $ 2 \g^2 Q^3$ and therefore $Q$ is supported for a sizeable duration of the inflationary period.  The evolution of the inflaton, $\tau_2$, as  of $\xi_f$, $\xi_h$ and the slow-roll parameters, $\epsilon, \epsilon_{\phi},\epsilon_B,\epsilon_E$ as well as $f_c$ are shown in FIGs. \ref{tau2KI2}-\ref{epsvBKI2}. 
Notice in particular that as we discussed in section \ref{Sec2}, $\xi_h\gtrsim \xi_Q$, where in the present example $\xi_h =\frac{m\,\dot b}{2\tau_4\dot \varphi}\sqrt{2\epsilon_\varphi}$ and one can check that the condition \eqref{xiseq} is satisfied. 

Notice also  that $\epsilon_B \sim \epsilon > \epsilon_{\varphi}$ in the early stages of inflation. This is a similar  but less dramatic situation to  that described in \cite{Fujita17} and is consistent as long as the scalar perturbations of the gauge field are very small relative to the inflationary perturbation and therefore the scalar power spectrum can be taken as $\frac{H^2}{8\pi\epsilon_{\varphi}}$, which we  discuss in section \ref{Sec5}.

We turn now to the inflationary predictions of the model dictated by the inflaton $\tau_2$. 
Let us start discussing the  scalar spectral index: 
\beq\label{ns}
n_s= 1 - 2\epsilon - \eta_{\varphi}\,,
\eeq
with 
$ \eta_{\varphi}=d\left(\ln \epsilon_{\varphi}\right)/dN$.  This form for $n_s$ arises under the assumption that the power spectrum is well-approximated by ${\cal P}_S = \frac{H^2}{8\pi^2\epsilon_{\varphi}}$ instead of ${\cal P}_S = \frac{H^2}{8\pi^2\epsilon}$. 
Since in the model described, $\epsilon \sim \epsilon_B \gg \epsilon_{\varphi}$ for much of inflation, this is an important distinction. 
This assumption is well-justified if the scalar power spectrum receives a negligible contribution from the gauge field scalar perturbations on super-horizon scales \cite{Fujita17} (see \ref{Sec5}) and is therefore the same power spectrum that would arise in this model if the gauge fields were ignored, i.e.~eq.~\eqref{ns}, while the expression for  $r$  is 
\beq
r\, =\, {\cal P}_T/{\cal P}_S = 16 \epsilon_{\varphi}\,.
\eeq
This model gives the following inflationary predictions, at 60 e-folds before the end of inflation: 
\bea
&& \epsilon_{\varphi}=2.80 \times 10^{-8}, \qquad n_s=0.964, \qquad r_b  = 4.48 \times 10^{-7}, \nonumber \\
&& \hskip1cm \Delta \varphi = 0.190 \Mp, \qquad H_{inf} \simeq 7\times 10^{-8}\Mp\,,
\eea
where $\Delta \varphi = \int_{N_*}^{N_e} \sqrt{2 \epsilon_{\varphi}}\,dN$ with $N_e$ the end of inflation and $N_*=N_e-60$, $n_s$ is the scalar spectral index, and $r= 16\, \epsilon_{\varphi}$ is the non-sourced estimate for the tensor-to-scalar ratio. As we can see, the predictions for $r$ are well below the observational target of $r\sim 10^{-3}$. 

We have shown that a viable background solution where the gauge field, $Q$, is sustained by its coupling to an axion is feasible in a K\"ahler inflation set-up including a suitable spectaor sector. In the next section, we turn to the perturbations of this system and consider whether this gauge field can act as a source for primordial gravitational waves.

\section{Cosmological  Perturbations } \label{Sec4}
In the previous sections we learned that it is possible to build a concrete scenario
within string theory that fulfils the  requirements for successful background dynamics for spectator chromonatural inflation realising our first two goals. As we discussed, this can be achieved by fixing two of the parameters, namely the magnetic flux, $m$, and the condensing group degree, $N$. Anticipating the requirement of a controllable backreaction from the tensor gauge fluctuations, we also fixed ${\tt g}$, which in turns fixes the last parameter, namely the wrapping number, $n$. 
 
\begin{figure}[H]
	\centering
	\includegraphics[width=0.7\textwidth]{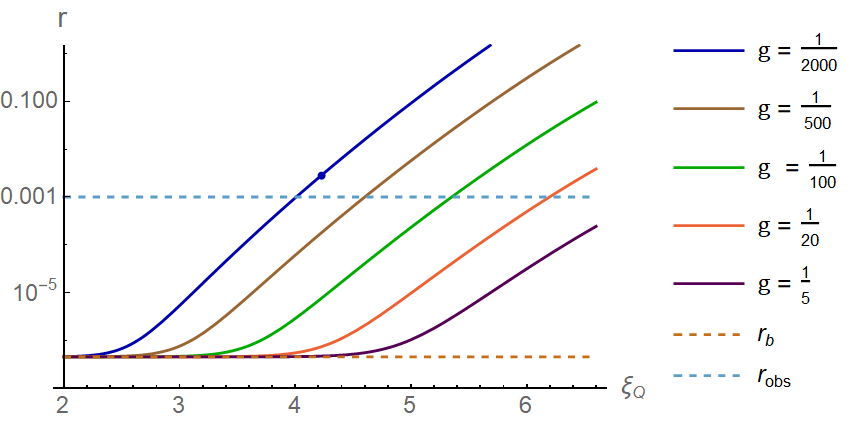}
	\caption{Analytic estimate of the sourced tensor-to-scalar ratio, $r$, against $\xi_Q = \frac{\g Q}{H}$ for different values of $\g$. The dashed lines correspond to the (non-sourced) background tensor-to-scalar ratio for our example (orange) and the observational cut-off $r \gtrsim 10^{-3}$ (blue), while the dot corresponds to the sourced tensor-to-scalar ratio for the example shown in this paper. }
	\label{Estima_r}
\end{figure}

\begin{figure}[H]
	\centering
	\includegraphics[width=0.55\textwidth]{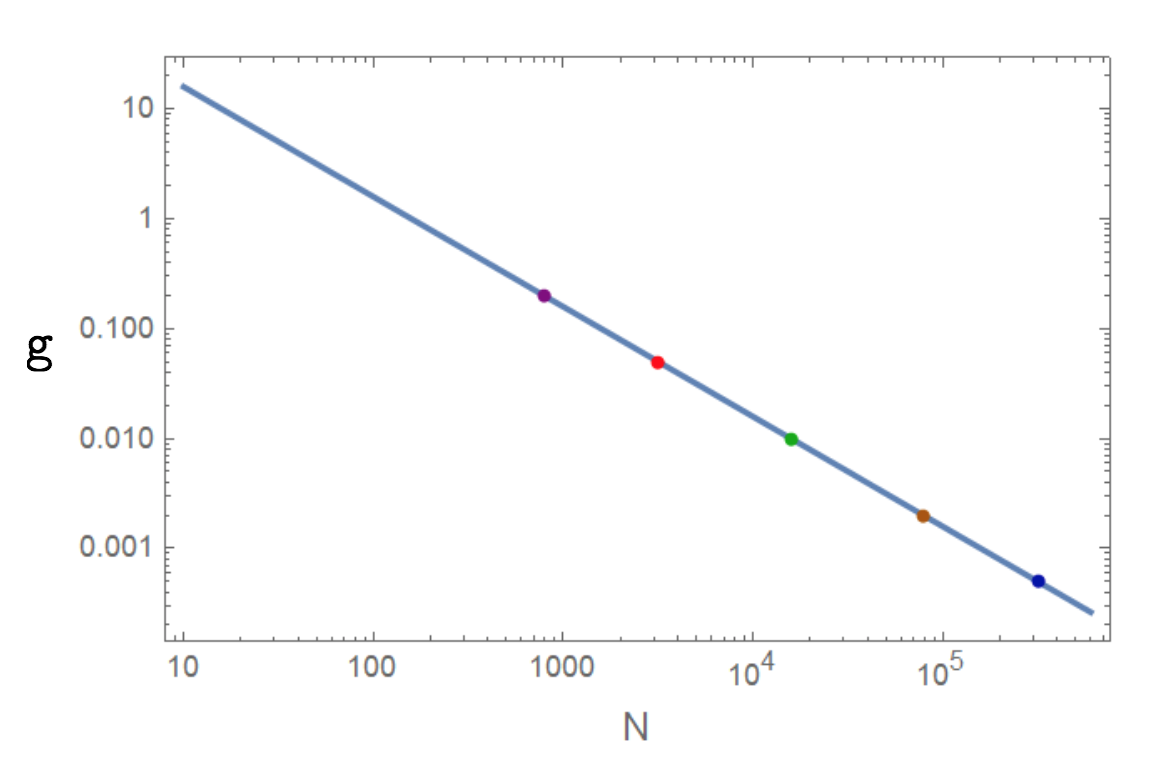}
	\caption{The effective coupling $\g$ as function of the gauge-group degree, $N$ for the value of $m$ and $\tilde a$ in \eqref{params2}. The colours of the dots are coordinated with figure \ref{Estima_r} and correspond to the different values of $\g$ that are used in that plot.}
	\label{gofN}
\end{figure}

\begin{figure}[H]
	\centering
	\includegraphics[width=0.7\textwidth]{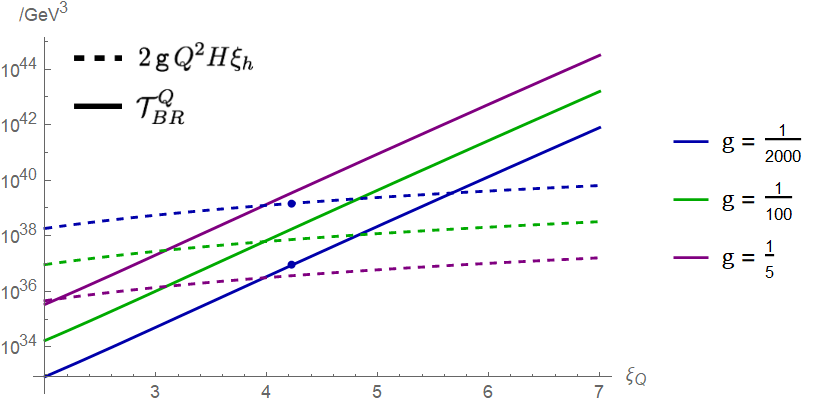}
	\caption{The backreaction of the tensor fluctuations on the equation of motion for $Q$, $\mathcal{T}^Q_{BR}$ (solid lines) along with the largest term in the equation of motion for $Q$, $2 \g Q^2 H \xi_h$ (dashed lines) plotted against $\xi_Q$ for different values of $\g$. Increasing $\xi_Q$ increases $\mathcal{T}^Q_{BR}$ much faster than $2 \g Q^2 H \xi_h$, but decreasing $\g$ reduces $\mathcal{T}^Q_{BR}$ whilst increasing $2 \g Q^2 H \xi_h$. The dots correspond to the values of $2 \g Q^2 H \xi_h$ (on dashed line) and $\mathcal{T}^Q_{BR}$ (on solid line) that are found in the example above 60 e-folds before the end of inflation.}
	\label{BackR}
\end{figure}

In this section we  consider the evolution cosmological perturbations to the  model described in section \ref{Sec3}. We are particularly interested to discuss the following aspects in more detail:
\begin{itemize}
\item {\it \underline{Amplification of primordial tensor fluctuations}}: we show that the primordial spectrum of tensor fluctuations  of K\"ahler inflation  can be amplified by a factor of order  $10^3$ when embedded in  our realisation of  chromonatural  inflation. The enhanced spectrum  reaches values that can  be probed by the future  generation of CMB polarisation experiments \cite{SPT3G,simons,cmb-s4,class,Lbird,pico}. The large enhancement corresponds to our second goal and requires fixing a second parameter, by requiring that the ``instantaneous" decay constant is large enough to sustain the gauge field for enough e-folds. This fixes a second available parameter, specifically the condensing group degree $N$ to be large.  The fact that the spectrum is chiral also means it is distinguishable from a higher-energy single-field model.
However, we shall explain that the tensor chirality  is not observable for the predicted
value of the parameters of our model.

\item {\it \underline{Backreaction of gauge fluctuations}:} 
 for the first time when discussing a string embedding of the SCNI model, we 
 perform a careful estimate of the backreaction of fluctuations following \cite{DFF}. We show  that  the backreaction of a potentially large amplitude of gauge field fluctuations 
  can be made subdominant in the evolution equations for background quantities. As anticipated, a small backreaction requires a small effective gauge coupling, which we defined in section \ref{Sec3} as (eq.~\eqref{geq2})
 \beq
 {\tt g} = \frac1{\sqrt{N n/2}},
 \eeq
 thus requiring fixing the  third parameter, namely the spectator D7-brane stack wrapping number, $n$. 

\end{itemize}
Before discussing the perturbations in detail, we show in figures \ref{Estima_r} - \ref{BackR} why the system is heavily constrained, and in particular why we require a very low value for $\g$ and what this means for $N$. In figure \ref{Estima_r} we plot the analytical estimate (see section \ref{sec:TP}) of the sourced tensor-to-scalar ratio as a function of the parameter $\xi_Q = \g Q/H$  for different values of  the effective gauge coupling, $\g$. The enhancement is exponentially sensitive to the value of $\xi_Q$. In figure \ref{gofN} we show  ${\tt g}$ as a function of the gauge group degree $N$ and in figure  \ref{BackR} we show an estimate of the backreaction (see section \ref{Sec6}) of the gauge field tensor perturbations on the equation of motion for $Q$ \eqref{GF1} as a function of  $\xi_Q$  for different values of  ${\tt g}$. In order to ensure that our approach of choosing the gauge field to be in the isotropic configuration, $A_i^A = \delta^A_i a Q$, is consistent, we require that the backreaction $\mathcal{T}^Q_{BR}$, be considerably smaller than the largest term in \eqref{GF1}, namely $2 \g Q^2 H \xi_h$. The backreaction also increases exponentially with $\xi_Q$. However, the lower $\g$, the smaller the backreaction. Since we require a large enhancement to uplift the tensor spectrum of K\"ahler moduli inflation, we need a relatively large value for $\xi_Q$ (in \cite{DFF}, $\xi_Q \sim 3.5$) and consequently a very small value for $\g$. 

\subsection{Set-up}\label{sec:setup}
Following \cite{DP}, the perturbations can be decomposed as:
\bea
&& \phi^a=\phi^a(t)+\delta \phi^a(t,x^i) \nonumber \\
&& A_0^A=a(t) \left(Y_A(t,x^i)+\partial_A Y(t,x^i)\right) \nonumber \\
&& A_i^A=a(t) \left[ (Q(t)+\delta Q(t,x^i)) \delta_{Ai}+\partial_i(M_A(t,x^i)+\partial_A M(t,x^i))  \right. \nonumber \\
&&\hskip6cm \left.
+\epsilon_{iAC}\left(U_C(t,x^i)+\partial_CU(t,x^i)\right)+t_{iA}(t,x^i)\right] \nonumber \\
&& g_{00}=-a^2(t)(1-2 \phi(t,x^i)) \nonumber \\
&& g_{0i}=a^2(t)(B_i(t,x^i)+\partial_iB(t,x^i)) \nonumber \\
&& g_{ij}=a^2(t)\left[(1+2 \psi(t,x^i))\delta_{ij}+2 \partial_i\partial_jE(t,x^i)+\partial_iE_j(t,x^i)+\partial_jE_i(t,x^i)+h_{ij}(t,x^i)\right] \nonumber \\
&&
\eea
Here, again, $A=1,2,3$ is the SU(2) index and $i=1,2,3$ is the spatial index. The tensor modes are the perturbations $t_{ij}$ and $h_{ij}$, on which we impose the transverse and traceless gauge: $\partial_ih_{ij}=\partial_it_{ij}=t_{ii}=h_{ii}=0$ which leaves us with 4 tensor perturbations. The vector modes are $Y_A,M_A,U_C,B_i,E_i$, which are also chosen to be transverse leaving 10 vector perturbations. The scalar modes contribute another 11 perturbations. However, the SU(2) gauge freedom allows us to set $U=U_i=0$, immediately removing a scalar and vector perturbation.

\subsection{Tensor perturbations}\label{sec:TP}
First we focus on the tensor perturbations, and discuss the amplification of primordial
tensor modes\footnote{We will use a different formalism for the metric when working with the scalars for convenience.}. Taking our wave-vector along the $z$-axis, $k=k_z$ and including only the remaining tensor perturbations, we can write the gauge fields and metric as:
\bea
&& A^1_{\mu}=a\,(0,Q+T_+,T_{\times},0)\, \nonumber \\
&& A^2_{\mu}=a\,(0,T_{\times},Q-T_+,0)\, , \nonumber \\
&& A^3_{\mu}=a\,(0,0,0,Q)
\eea
and 
\beq
g_{\mu\nu}=a^2\begin{pmatrix}
	-1 && 0 && 0 && 0 \\
	0 && 1+h_+ && h_{\times} && 0 \\
	0 && h_{\times} && 1-h_+ && 0 \\
	0 && 0 && 0 && 1
\end{pmatrix} \, 
\eeq
where $T_+,T_{\times}$ and $h_+,h_{\times}$  are the transverse tensor perturbations of the gauge fields and metric, respectively.
To quadratic order the  inverse metric is:
\beq
g^{\mu\nu}=a^{-2}\begin{pmatrix}
	-1 && 0 && 0 && 0 \\
	0 && 1- h_++h_+^2+h_{\times}^2 && -h_{\times} && 0 \\
	0 && -h_{\times} && 1+h_++h_+^2+h_{\times}^2 && 0 \\
	0 && 0 && 0 && 1
\end{pmatrix} \, .
\eeq

\subsubsection{Analytic approximation}\label{anaapprox}

After finding the tensor action to 2nd order in perturbations, the linearised equations of motion for the tensor modes to leading order in slow-roll can be found, and these are given below. We use  standard notation used in the literature (as e.g.~in \cite{DFF}), whereby we split the tensor modes into left and right-moving modes as: 
$$\psi_{L,R}=\frac{a\Mp}{2}(h_+\pm i h_{\times}) \qquad {\rm and}\qquad  t_{L,R}=a(T_+\pm iT_{\times}).$$ 
By defining
\beq
\label{defOFx}
x \,=\,\frac{k}{a H}
\eeq
and 
applying a first order slow-roll expansion\footnote{Here by first order, we mean objects proportional to $\sqrt{\epsilon_i}$, $i=\varphi,E,B$.}, we arrive at the equations of motion:
\beq 
\partial_x^2 \psi_{R,L}+\left(1-\frac{2}{x^2}\right)\psi_{R,L}=\frac{2  \sqrt{f \epsilon_E}}{x}\partial_xt_{R,L}+\frac{2\sqrt{f\epsilon_B}}{x^2}\left(\xi_Q\mp x\right)t_R \label{psirl}
\eeq
\bea
&&\partial_x^2 t_{R,L}+\left[1+\frac{2}{x^2}(\xi_Q\xi_h\mp x(\xi_Q+\xi_h))\right]t_{R,L}-\frac{2\xi_{f}}{x}\partial_x t_{R,L}\nonumber \\ 
&&=-\frac{2 \sqrt{\epsilon_E/f}}{x}\partial_x\psi_{R,L}+\frac{2}{x^2}\left[(\xi_Q\mp x)\sqrt{\epsilon_B/f}+(1+2\xi_{f})\sqrt{\epsilon_E/f}\right]\psi_{R,L} \label{trl}
\eea
where 
\bea
\sqrt{\epsilon_E}=\frac{\sqrt{f}}{\Mp}\left(Q-x \del_x Q\right) \,, \qquad   \sqrt{\epsilon_B}=\frac{\sqrt{f} \g\, Q^2}{H \Mp}\,,
\eea
and recall that 
\bea
\xi_Q=\frac{\g Q}{H} \,, \qquad \xi_h=\frac{m\, \dot{b}}{2f H} \,, \qquad
\xi_{f}= \frac{ \dot{\tau}_4}{2f H}\, .
\eea
Before discussing our numerical results, we can make some progress analytically as in \cite{DFF}. First, we note that only the right-helicity mode of the gravity waves can be enhanced -- this is because only the right-helicity mode of the gauge tensor perturbation, $t_R$, has a growing mode at any time period. The relevant term in \eqref{trl} is given by $\mp \frac2x \left(\xi_Q+\xi_h\right)t_{R,L}$, which can only lead to a growing solution in the right-helicity mode, $t_R$, due to the minus sign\footnote{Due to the $1/x$ factor, and $\xi_Q,\xi_h>1$, this term dominates over the other two terms proportional to $t_{R,L}$ when $x \sim \mathcal{O}(1)$, and $t_R$ becomes a growing mode. When $x\gg 1$, the term $t_{R,L}$ dominates, and both $t_{R,L}$ decay; and then when $x \ll 1$, the term $\frac{2}{x^2}\xi_Q\xi_h t_{R,L}$ dominates and both solutions decay again. The terms on the RHS of \eqref{trl} are negligible and do not affect the evolution of $t_{R,L}.$}. For this reason, we will only consider the right-helicity mode, $t_R$, in our discussion as this is the one relevant to enhancing the gravitational wave spectrum.  Proceeding to find an analytic approximation, we start by solving the right-helicity homogeneous equation of \eqref{trl}, which has solutions (assuming $\xi_Q$, $\xi_h$ and $\xi_f$ are constant)
\beq\label{tsol}
t_{R}=\frac{1}{\sqrt{2k}}i^{\beta}x^{\xi_f}W_{ \beta,\alpha}(-2i x)
\eeq
where $W_{k,m}(z)$ is the Whittaker function, $\beta=-i(\xi_Q+\xi_h)$ and $\alpha=\frac12 \sqrt{-8\xi_Q \xi_h+(1+2\xi_f)^2}$. The solution is normalised by $t_{R}\rightarrow (2k)^{-1/2}(2x)^{\beta}e^{ix}x^{\xi_f}$, $x \to \infty$, which is the equivalent condition to \cite{DFF} (set $\xi_f = 0$).

\begin{figure}[H]
	\centering
	\includegraphics[width=0.4\textwidth]{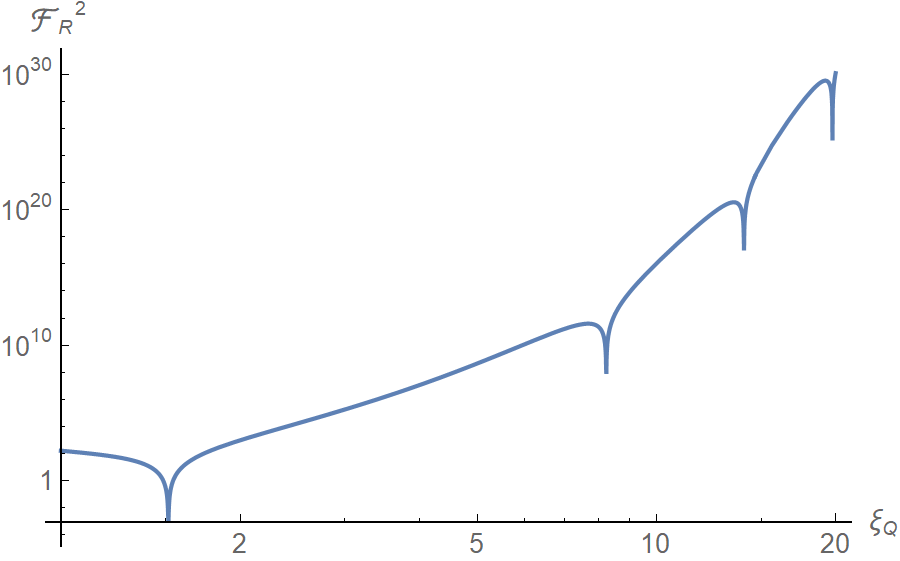}\qquad \includegraphics[width=0.4\textwidth]{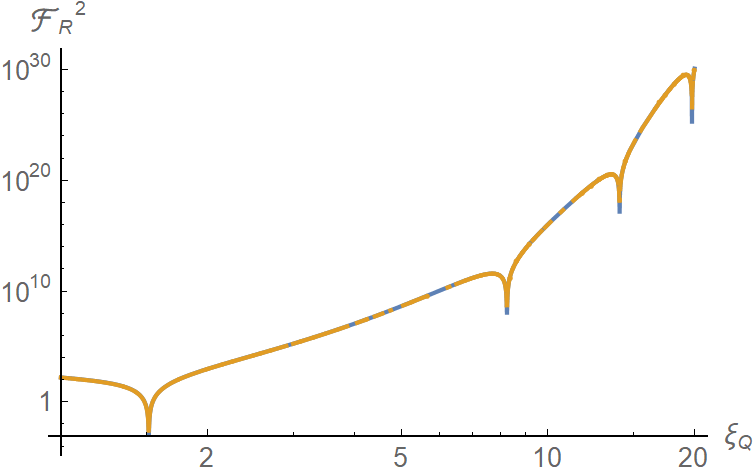}
	\caption{The amplification factor, $\mathcal{F}^2$, of the tensor power spectrum for right-helicity modes in our analytic estimate plotted against the effective mass of the gauge field, $\xi_Q$. The left figure uses $\xi_f=7.85\times 10^{-3}$ and the right figure shows two cases, $\xi_f=7.85\times 10^{-3}$ and $\xi_f=0$ demonstrating that $\xi_f$ is too small to have any effect on the tensor perturbations. }
	\label{Fsquared}
\end{figure}

We then wish to find the Green function for \eqref{psirl} for which we use the following formula 
\beq
G(x,x')=\frac{\nu_1(x)\nu_2(x')-\nu_1(x')\nu_2(x)}{\nu_1^{\, \prime}(x')\nu_2(x')-\nu_1(x')\nu_2^{\, \prime}(x')}
\eeq
where, $\nu_1$ and $\nu_2$ are the homogeneous solutions to the equation you wish to solve. In our case, \eqref{psirl} has homogeneous solutions:
\beq
\nu_1(x)=e^{ix}\left(1+\frac ix\right)
\eeq
\beq
\nu_2(x)=e^{-ix}\left(1-\frac ix \right) \, .
\eeq 
With this, our Green's function is
\beq
G(x,x')=\frac{(x'-x)\,\text{cos}(x'-x)-(1+xx')\,\text{sin}(x'-x)}{xx'} \, .
\eeq
We now need to perform the integral
\beq
\int dx'\left\{\frac{2 \sqrt{f\epsilon_E}}{x'}\partial_{x'}t_{R,L}+\frac{2 \sqrt{f\epsilon_B}}{(x')^2}(\xi_Q - x')\right\}G(x,x') \, .
\eeq
An analytic estimate for this integral can be found by first performing an indefinite integral over $x'$, then by taking two limits: first $x'\rightarrow \infty$ (the dominant contribution to this integral will be in the sub-horizon limit before the gauge field decays).
Then in the super-horizon limit
\beq
\lim_{x\to\, 0}\psi_{R}=\frac{1}{\sqrt{2k}x}\left\{\mathcal{F}_{E}^{R}\sqrt{f \epsilon_E}+\mathcal{F}_B^{R}\sqrt{f \epsilon_B}\right\}\, .
\eeq
The sourced right-helicity tensor power spectrum is given by:
\beq \label{tensorpower}
\mathcal{P}_h^{R,s}(k)=\frac{H^2}{\pi^2\Mp^2}\bigl \lvert \sqrt{2k} x \lim_{x\to0}\psi_{R}\bigr\rvert^2= f \frac{\epsilon_B H^2}{\pi^2 \Mp^2}\mathcal{F}_{R}^2
\eeq
where $\mathcal{F}^2_{R}\equiv \bigl\lvert \mathcal{F}_B^{R}+\sqrt{\epsilon_E/\epsilon_B}\mathcal{F}_E^{R} \bigr\rvert^2$ is a measure of the amplification.
$\mathcal{F}_E^{R}$ and $\mathcal{F}_B^{R}$ are given by

\bea
&&\mathcal{F}_E^R= \nonumber \\
&&-\frac{i\, 2^{4-\text{$\xi_f$}}}{256} e^{\frac{1}{2} i \pi  (\beta +\text{$\xi_f$})}\, \Gamma \left(-\alpha +\text{$\xi_f$}-\frac{3}{2}\right) \Gamma \left(\alpha +\text{$\xi_f$}-\frac{3}{2}\right)\bigg\{\frac{\Gamma (-\beta -\text{$\xi_f$})}{\Gamma \left(-\alpha -\beta +\frac{1}{2}\right) \Gamma \left(\alpha -\beta +\frac{1}{2}\right)}
 \nonumber \\
&&
\left[ \right.
 \left.  -16 \alpha ^4  +8 \alpha ^2 \left(8 \beta +4 \text{$\xi_f$}^2+5\right)-8 \left((8 \beta +3) \text{$\xi_f$}^2+8 \beta  \text{$\xi_f$}+2 \beta  (8 \beta +1)+2 \text{$\xi_f$}^4\right)+32 \text{$\xi_f $}-9\right]  \nonumber \\
&&+\frac{1}{\Gamma (-\beta +\text{$\xi_f$}+1)}\left[16 \alpha ^4+8 \alpha ^2 \left(8 \beta -4 \text{$\xi_f$}^2-5\right)+128 \beta ^2-16 \beta  (2 \text{$\xi_f$}+1)^2\right. \nonumber \\
&& \left. +(1-2 \text{$\xi_f$})^2 (4\, \text{$\xi_f$} \,(\text{$\xi_f$}+1)+9)\right]\bigg\}\,,
\eea \label{FE}
\bea
&&\mathcal{F}_B^R = \nonumber \\
&&\frac{2^{4-\xi _f}}{256} e^{\frac{1}{2} i \pi  \left(\beta +\xi _f\right)} \Gamma \left(-\alpha +\xi _f-\frac{3}{2}\right) \Gamma \left(\alpha +\xi _f-\frac{3}{2}\right) \bigg\{ \frac{\Gamma \left(-\beta -\xi _f\right)}{\Gamma \left(-\alpha -\beta +\frac{1}{2}\right) \Gamma \left(\alpha -\beta +\frac{1}{2}\right)}
 \nonumber \\
&&
\left[ \right.
\left. 
8i\xi_Q(\beta +\xi _f)\left(1-4 \alpha ^2+8 \beta +4 \left(\xi _f-1\right) \xi _f\right) \right. \nonumber \\
&& \left. -\left(4 \alpha ^2-4 \left(\xi _f-3\right) \xi _f-9\right) \left(4 \alpha ^2-8 \beta -4 \xi _f \left(\xi _f+1\right)-1\right)\right] \nonumber \\
&&-\frac{1}{\Gamma \left(-\beta +\xi _f+1\right)}\left[8 i \xi_Q \left(\beta -\xi _f\right) \left(4 \alpha ^2+8 \beta -4 \left(\xi _f-1\right) \xi _f-1\right) \right. \nonumber \\
&& \left.+\left(4 \alpha ^2-4 \left(\xi _f-3\right) \xi _f-9\right) \left(4 \alpha ^2+8 \beta -4 \xi _f \left(\xi _f+1\right)-1\right)\right]\bigg\}\,.
\eea \label{FB}
 We saw in section \ref{Sec2} that in the slow-roll approximation, the relation \eqref{xiseq}, $\xi_h \simeq \xi_Q+\xi_Q^{-1}+\xi_f/\xi_Q$, among $\xi_h, \xi_Q$ and $\xi_f$ holds and moreover, $\xi_f\ll1$.  
 Fixing $\xi_f=7.85\times 10^{-3}$, the value it takes 60 e-folds before the end of inflation in the model discussed in section \ref{Sec3}, we plot the amplification factor $\mathcal{F}^2$ for the right mode in FIG. \ref{Fsquared} and  compare the  amplification factor for $\xi_f=7.85\times 10^{-3}$ and $\xi_f=0$. As we see from the plot, there is a negligible difference, indicating  that this coupling  does not  affect the tensor spectrum.

We define the total tensor spectrum as $r=r_b+r_{s}$ where $r_b$ is the background inflationary tensor spectrum $r_b=(2H^2/\pi^2)/{\cal P}_S$  and $r_{s}={\cal P}^s_h/{\cal P}_S$ with ${\cal P}_S=2.1 \times 10^{-9}$ where ${\cal P}^s_h$ is the sourced part of the tensor power spectrum and is calculated through \eqref{tensorpower}. For this example, 60 e-folds before the end of inflation we have $r_b=4.48 \times 10^{-7}$ and $r_{s}=2.86 \times 10^{-3}$ leading to an overall estimate for the tensor-to-scalar ratio of $r=2.86 \times 10^{-3}$. 

It is interesting to stress that the resulting tensor spectrum is fully chiral, since only the right-helicity tensor modes get amplified. On the other hand, the resulting tensor-to-scalar ratio is at least  one order of magnitude too small for detecting chirality 
 by cross-correlating $T$, $E$ and $B$ spectra with future CMB experiments -- see e.g. \cite{Gluscevic:2010vv} for a detailed analysis.

\subsubsection{Numerical results}
\begin{figure}[H]
	\centering
	\includegraphics[width=0.7\textwidth]{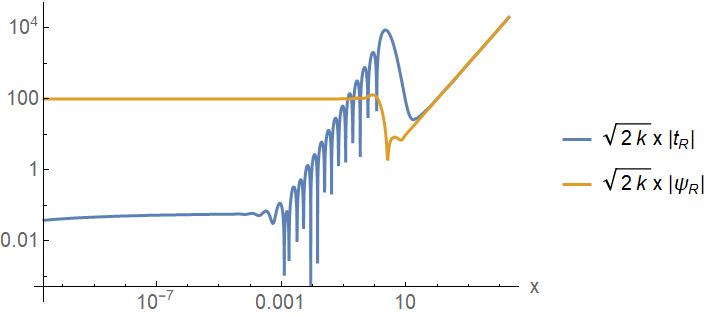}
	\caption{The evolution of the (right-helicity) tensor modes for the gauge field, $t_R$, and the gravity sector, $\psi_R$,  plotted against $x=k/aH$.}
	\label{psit}
\end{figure}

We  now discuss the full numerical solution for the tensor perturbations of the model discussed in section \ref{Sec3}.  We consider the full equations of motion without employing the slow-roll approximation and normalise our solutions in the Bunch-Davies form $t_R(x_{in})=\psi_R(x_{in})=1/\sqrt{2k}$, $t_R^{\, \, \prime}(x_{in})=\psi_R^{\, \, \prime}(x_{in})=i/\sqrt{2k}$ where $x_{in}$ should be some relatively large number that we take to be $x_{in} = 2\times10^4$, numerically approximating infinity, and $k=k_*=0.05\, \text{Mpc}^{-1}$. The evolutions of $t_R$ and $\psi_R$  are plotted in FIG.\ref{psit}. With no enhancement to the gravity sector, $\lvert\sqrt{2k}\, x\, \psi\rvert\rightarrow 1$ at super-horizon scales, $x<1$, but as we can see $\psi_R$ freezes out at super-horizon scales with an enhanced value due to a transient instability experienced by $t_R$ just before horizon-crossing even as $t_R$ decays. The freeze-out value of $ \lvert\sqrt{k}\,x\,\psi_R\rvert^2$ is the amplification factor for the tensor power spectrum. Evaluating the tensor-to-scalar ratio, $r={\cal P}_h/{\cal P}_S$ with this freeze-out value leads to a similar enhancement to that predicted by our analytic estimate above, giving an enhancement of 

\beq
r_b=4.48 \times 10^{-7} \, \longrightarrow r=2.29 \times 10^{-3}\,
\eeq
 an amplification of $5113$ for the example shown in section \ref{Sec3}. The value of $r=2.29 \times 10^{-3}$ is slightly smaller than our analytic estimate ($2.86 \times 10^{-3}$) but still large enough to be potentially observable at next generation detectors such as CMB-S4 \cite{cmb-s4} and many others \cite{SPT3G,Lbird,class,simons,pico}. We define the tensor power spectrum as\footnote{This is not the same as \eqref{tensorpower}, which accounted only for the sourced contribution to the tensor power spectrum. The following is defined through the full numerical solution and is therefore the full tensor power spectrum.}:
\beq
{\cal P}_h=\frac{H^2}{\pi^2\Mp^2}\bigl \lvert \sqrt{2k}\, x\,\psi_R \bigr \rvert^2
\eeq
and evaluate this with the freeze-out value ($x\ll1$) for $\bigl \lvert \sqrt{2k}\, x\,\psi_R \bigr \rvert^2$, with horizon-crossing, $x=1$, taken to be 60 e-folds before the end of inflation.

\noindent Comparing this to the example given in \cite{DFF} where the freeze-out value of $\sqrt{2k} x \psi_R \lesssim 10$ (FIG. 4), we see that the amplification factor in our model is much larger. This is of course necessary because we wish to amplify the tensor-to-scalar ratio to observable values $r \gtrsim 10^{-3}$. In the example given in \cite{DFF}, $r_b \sim 10^{-3}$ whereas in our example, $r_b \sim 10^{-7}$ meaning we require a much greater enhancement. This does not come for free and the larger enhancement leads to larger backreaction (see section \ref{Sec6}), which can be compensated for by reducing the value of $\g$, and in \cite{DFF}, $\g = 1.11 \times 10^{-2}$, compared to our value of $\g = 5 \times 10^{-4}$. 

We have shown both with an analytic approximation and using a full numerical solution that it is possible to greatly amplify the tensor power spectrum of K\"ahler moduli inflation. In the case of the full numerical solution, we have shown that it is possible to amplify the tensor power to observable values, $r \gtrsim 10^{-3}$. In the next two sections, we will demonstrate that first the scalar perturbations are under control, and second that this great enhancement does not lead to excessive backreaction.

\subsection{Scalar perturbations} \label{Sec5}

We  now discuss the scalar perturbations in this system. As we mentioned above, we  use, as a matter of convenience, a different formalism for the metric to the one used above for the tensor perturbations. Using this formalism, we will show that the scalar metric perturbations have no impact at linear order, and can therefore be neglected in the evolution. Our starting point is the ADM formalism as described for generalised multi-field inflation in \cite{Langlois:2008mn}. In the ADM formalism, the metric is taken to be:
\beq
ds^2=-N^2dt^2+ h_{ij}(dx^i+N^i\,dt)(dx^j+N^j\,dt)
\eeq
where $N$ and $N^i$ are the lapse and shift functions, respectively.
With this choice, our full action, \eqref{action2}, can be written in the form:
\beq
S=\frac12 \int dtd^3x\sqrt{h}N\left(R^{(3)}+2P\right)+\frac12\int dtd^3x\frac{\sqrt{h}}{N}\left(E_{ij}E^{ij}-E^2\right)\,,
\eeq
where $R^{(3)}$ is the Ricci scalar calculated with the spatial metric, $h_{ij}$, whose determinant is $h=\text{det}\left(h_{ij}\right)$. $E_{ij}$ is the symmetric extrinsic curvature tensor given by:
\beq
E_{ij}=\frac12 \dot{h_{ij}}-\frac12\nabla^{(3)}_jN_i-\frac12\nabla^{(3)}_iN_j
\eeq
and $E$ is its trace, $E=h^{ij}E_{ij}$. Finally, $P$, is the matter sector and for our system is:
\beq
P=X-V-\frac {f(\phi^a)}4 F_{\mu\nu}^AF^{A\,\mu\nu}+\frac {h(\phi^a)}4
\tilde{F}^A_{\mu\nu}F^{A\,\mu\nu}
\eeq
where $X$ is the kinetic term and can be decomposed as:
\beq
X=-\frac12 \gamma_{a b}\partial_{\mu}\phi^a \partial^{\mu}\phi^b=\frac{1}{2N^2}\gamma_{a b}v^av^b-\frac{\gamma_{a b}}{2}
h^{ij}\partial_i\phi^a\partial_j\phi^b
\eeq
with $v^a=\dot{\phi}^a-N^j\partial_j\phi^a$.
Variation with respect to the lapse, $N$, leads to the energy constraint:
\bea
&& R^{(3)}-2V-\gamma_{a b}h^{ij}\partial_i\phi^a\partial_j\phi^b-\frac{1}{N^2}\left\{E_{ij}E^{ij}-E^2+\gamma_{a b}v^av^b\right. \nonumber \\ 
&&\left.+f\left(h^{ij}F_{0i}^AF_{0j}^A-h^{jk}N^iF_{0j}^AF_{ik}^A+h^{ij}N^kN^lF_{ik}^AF_{jl}^A+h^{ij}N^kF_{0i}^A F_{jk}^A\right)\right\}=0 \, \label{encon}
\eea
and variation with respect to the shift, $N^i$, leads the momentum constraint:
\beq
\nabla_j^{(3)}\left(\frac{1}{N}\left(E^j_i-E\delta^j_i\right)\right)N=\gamma_{a b}v^a\partial_i\phi^b+f\,h^{jk}\left(N^lF_{kl}^AF_{ij}^A+F_{0j}^AF_{ik}^A\right)\, . \label{momcon}
\eeq
We now wish to linearise the system in scalar perturbations. We choose the spatially flat gauge to set $h_{ij}=a(t)^2\delta_{ij}$ and define the scalar perturbed lapse and shift function as:
\beq
N=1+\alpha\,, \qquad N_i=\partial_i\beta
\eeq
where $\alpha$ and $\beta$ are linear perturbations.
As defined in subsection \ref{sec:setup}, we decompose our fields in the following way (considering only scalar perturbations):
\bea
&&\phi^a=\phi^a(t)+\delta \phi^a(t,x^i) \nonumber \\
&& A_0^A=a(t)\,\partial_A Y(t,x^i) \nonumber \\
&& A_i^A=a(t)\left[\left(Q(t)+\delta Q(t,x^i)\right)\delta_{Ai}+\partial_i\partial_AM(t,x^i)\right]\,.
\eea
For our  example in section \ref{Sec3}, $\delta \phi^a=\{\delta \tau_2,\delta \tau_4, \delta b\}$. Including $\alpha$ and $\beta$, we therefore have 8 remaining perturbations. However, the perturbations $Y,\alpha, \beta$ are all non-dynamical (no time derivatives of, e.g, $Y$, appear in its equation of motion). The equations of motion for these three perturbations can therefore be used as constraints. From the energy constraint \eqref{encon}, we get the following equation for $\beta$:
\bse
4H\partial^2\beta+12\, a^2 H^2 \alpha= 
2a H  \sqrt{f\epsilon_E}\,\partial^2Y-6a^2  H \sqrt{f\epsilon_E}\,(H\delta Q+\dot{\delta Q})-2a^2\,V_{,a}\,\delta \phi^a
-2a^2  H \sqrt{f\epsilon_E}\, \partial^2(HM+\dot{M})-2a^2\gamma_{a b}\dot{\phi^a}\mathcal{D}_t \delta \phi^b \label{bet}
\ese
and from the momentum constraint \eqref{momcon}, we get the $\alpha$ equation:
\beq\label{alp}
\alpha=\frac{1}{2H}\left(\gamma_{a b}\,\dot{\phi^a}\delta\phi^b +2  H \sqrt{f\epsilon_E}\,\delta Q+2H\xi_Q\sqrt{f\epsilon_B}\,Y\right) \, ,
\eeq
where $\mathcal{D}_t\delta \phi^a=\dot{\delta \phi^a}+\Gamma^a_{bc}\,\dot{\phi^b}\delta \phi^c$, $\partial^2=\partial_i \partial^i$, and  we have performed a systematic slow-roll expansion up to first order in $\sqrt{\epsilon}$ as was described for the tensor perturbations. By expanding the action up to second order in the scalar perturbations, and substituting the values  for $\alpha$ and $\beta$ found from \eqref{bet} and \eqref{alp}, we find equations for the remaining perturbations, $Y,\delta \phi^a,\delta Q,M$. The constraint equation for $Y$ can then be found after moving into momentum space and choosing the wave-vector to be purely on the $z$-axis, $k=k_z$. $Y$ can be found from this equation then substituted into the Fourier-space equations of motion for $\delta \phi^a,\delta Q,M$. Whether we set the metric perturbations $\alpha$ and $\beta$ to zero or not, the algebra is extremely involved, and we will therefore only summarise our results here. Before attempting to numerically solve the equations for $\delta \phi^a,\delta Q,M$, we change to $x=k/aH$ coordinates, and by redefining the fields as in \cite{DFF,DP}:
\bea\label{scalarperts}
&& \delta \tau_2=\frac{\Delta_{\tau_2}}{a}\,, \qquad \delta \tau_4 =\frac{\Delta_{\tau_4}}{a}\,, \qquad  \delta b = \frac{\Delta_{b}}{a} \nonumber \\
&& \delta Q=\frac{\Delta_1}{\sqrt{2}a}\, , \qquad M=\frac{a{\tt{g}}Q\Delta_1+\sqrt{k^2+2a^2{\tt{g}}^2Q^2}\Delta_2}{\sqrt{2}{\tt{g}}a^2k^2Q}
\eea
we can remove $a$ and $k$ from the equations of motion, using $k=x a H$. The initial conditions for the scalar field perturbations have the multi-field inflation form \cite{Weinberg:2008zzc}:
\beq
\Delta^a(x_{in})=\frac{-x H \frac{d\phi^a}{dx}}{\sqrt{2k} \dot{\varphi}}, \qquad \frac{d\Delta^a}{dx}(x_{in}) = \frac{-i x H \frac{d\phi^a}{dx}}{\sqrt{2k} \dot{\varphi}}
\eeq
where $\Delta^a =(\Delta_{\tau_2},\Delta_{\tau_4},\Delta_{b})$ and $\dot{\varphi}^2=\gamma_{ab}\dot{\phi^a}\dot{\phi^b}= x^2 H^2 \,\gamma_{ab}\frac{d\phi^a}{dx} \frac{d\phi^b}{dx}$ while the scalar gauge field perturbations have the standard Bunch-Davies initial conditions:
\beq
\Delta_i(x_{in})=\frac{1}{\sqrt{2k}}, \qquad \frac{d \Delta_i}{dx}(x_{in})=\frac{i}{\sqrt{2k}}
\eeq
where $\Delta_i=(\Delta_1,\Delta_2)$, $k=k_*=0.05\, \text{Mpc}^{-1}$ is the pivot scale and as before $x_{in}=2 \times 10^4$.

\begin{figure}[H]
	\centering
	\includegraphics[width=0.45\textwidth]{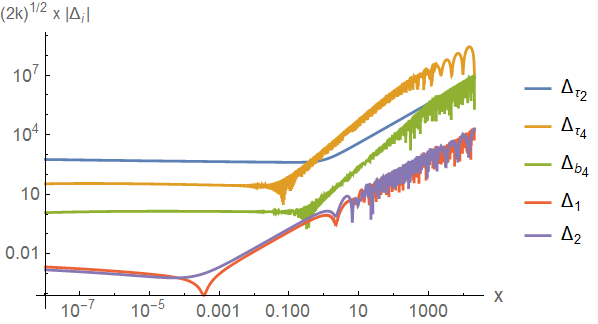}\qquad \includegraphics[width=0.45\textwidth]{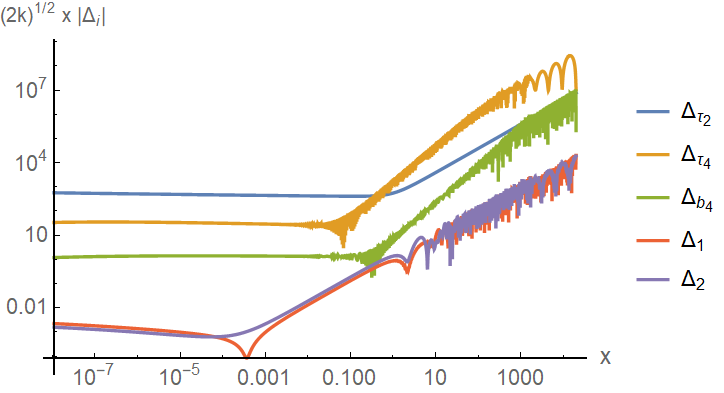}
	\caption{The evolution of the scalar perturbations defined in \eqref{scalarperts} with (left figure) and without (right figure) the inclusion of the metric perturbations $\alpha,\beta$ found through \eqref{alp}, \eqref{bet} for the example in section \ref{Sec3} plotted against $x=k/aH$. As we can see, the metric perturbations have a negligible effect on the evolution and can safely be set to zero.}
	\label{scalar perts}
\end{figure}

We  first demonstrate that the inclusion of the scalar metric perturbations, $\alpha, \beta$, in the equations of motion for scalar perturbations \eqref{scalarperts} has no effect on the evolution as these contribute negligibly. The complexity of the equations including the contribution from the metric perturbations is so high that a full evolution of the system is beyond the scope of this paper. Instead, we  show that over a shorter evolution\footnote{The smaller the final value of $x$, the closer to the end of inflation.}, there is no difference in the system's evolution with or without the inclusion of the metric perturbations. We then do a full evolution without the metric perturbations to include the decays of all the fields. As can be seen in FIG. \ref{scalar perts}, the metric perturbations have a negligible effect on the evolution of the scalar perturbations and can therefore be set to zero.

\begin{figure}[H]
	\centering
	\includegraphics[width=0.7\textwidth]{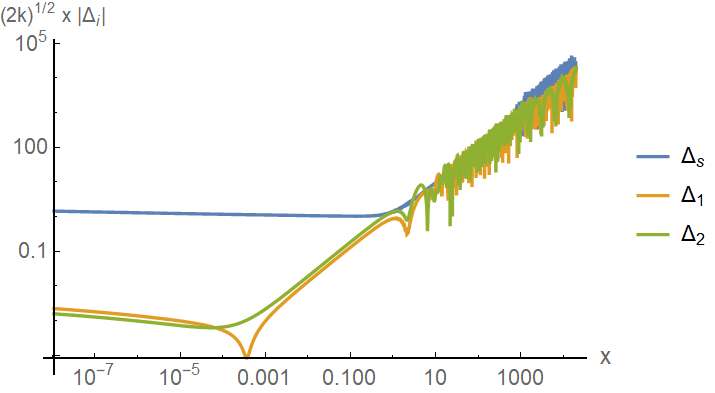}
	\caption{The evolution of the scalar perturbations defined in \eqref{scalarperts} and \eqref{mfs} plotted against $x=k/aH$. After horizon-crossing, $x<1$, the scalar perturbation associated with the scalar fields, $\Delta_s$ is consistently much larger than the scalars associated with the gauge field, $\Delta_1,\Delta_2$.}
	\label{scalar 1 pert}
\end{figure}

\begin{figure}[H]
	\centering
	\includegraphics[width=0.45\textwidth]{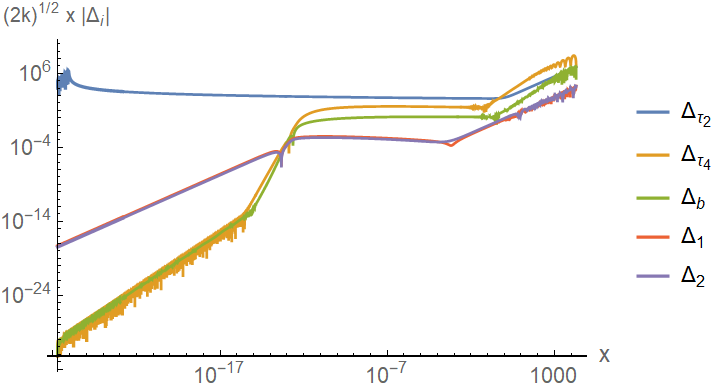}\qquad \includegraphics[width=0.45\textwidth]{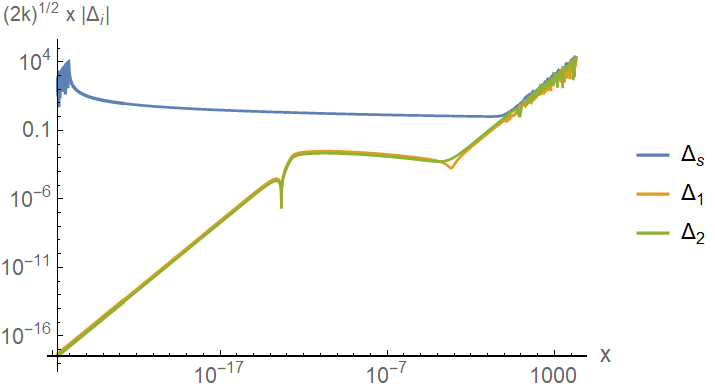}
	\caption{The full evolution of the scalar perturbations defined in \eqref{scalarperts} and \eqref{mfs} plotted against $x=k/aH$ including the decays of all the perturbations which take place concurrently with the decays of the corresponding background fields. }
	\label{scalar perts long}
\end{figure}

In order to see that the perturbations are well behaved, we define the (tangential) multi-field scalar perturbation

\beq\label{mfs}
\delta_s = \frac{\gamma_{ab}\,\dot{\phi^a}\, \delta \phi^b}{\dot{\varphi}} = \frac{\Delta_s}{a} =\frac{\gamma_{ab}\,\dot{\phi^a}\, \Delta^b}{a \,\dot{\varphi}}
\eeq
with $\Delta^a=(\Delta_{\tau_2},\Delta_{\tau_4},\Delta_{b})$. In the standard multi-field inflation case, the combination $\bigl \lvert \sqrt{2k} x \Delta_s \bigr \rvert$ should be $\mathcal{O}(1)$\footnote{In the perfectly massless case, when the potential is perfectly flat, $\bigl \lvert \sqrt{2k} x \Delta_s \bigr \rvert = 1$ after horizon-crossing. Therefore a value close to $1$ is expected during slow-roll inflation. When the slow-roll approximation breaks down, as happens when the inflaton nears its minimum, the perturbation will grow before decaying as the background inflaton settles to its minimum (see  FIG.~\ref{scalar perts long}).}  after horizon-crossing, $x<1$ before the background fields decay and  lead to a decay in the scalar perturbations. In FIG.~\ref{scalar 1 pert}, $\Delta_s$ is plotted with the gauge field scalar  perturbations, $\Delta_1,\Delta_2$. As can be seen, $\Delta_s$ freezes out ($x<1$) with $\bigl \lvert \sqrt{2k} x \Delta_s \bigr \rvert \sim \mathcal{O}(1)\gg \bigl \lvert \sqrt{2k} x \Delta_1 \bigr \rvert,\bigl \lvert \sqrt{2k} x \Delta_2 \bigr \rvert$ suggesting the gauge field has a negligible effect on the scalar power spectrum. The full evolution including the decays of all the perturbations is shown in FIG.~\ref{scalar perts long}. 

We have shown that the metric scalar perturbations' contribution is negligible and that the scalar perturbations to the gauge field are very small relative to the tangential inflationary perturbation. Our assumption is that the scalar power spectrum can be split into a contribution from the gauge field and the standard inflationary part. With the gauge field perturbations sub-dominant, this means that the power spectrum is  well-approximated by ${\cal P}_S = \frac{H^2}{8 \pi^2 \epsilon_{\varphi}}$ and that therefore the inflationary predictions of both ${\cal P}_S$ and $n_s$ are not spoiled by the presence of the spectator gauge field.

\subsection{Backreaction} \label{Sec6}

\begin{figure}[H]
	\centering
	\includegraphics[width=0.7\textwidth]{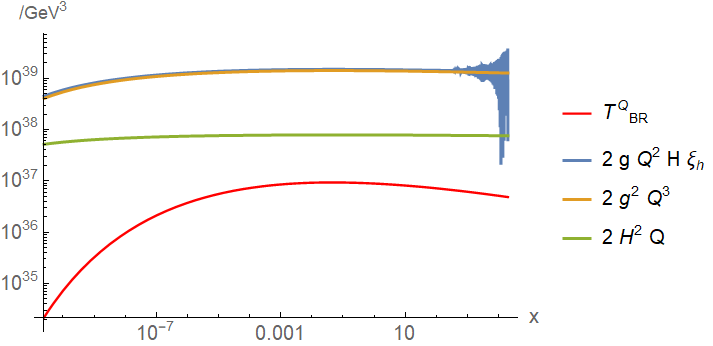}
	\caption{The evolution of the leading terms in the equation of motion for $Q$ \eqref{Q full}, $2 \g\,Q^2\,H\,\xi_h$ (blue), $2 \g^2\,Q^3$ (orange), $2H^2\,Q$ (green) and the backreaction induced by the gauge tensor perturbation, ${\cal T}^Q_{BR}$ (red) for the example shown in section \ref{Sec3} plotted against $x=k/aH$.}
	\label{backreaction}
\end{figure}

In order to verify the consistency of this model, particularly the assumption that at the background level, the gauge field, $A_i^A$, can be taken in the isotropic form, $A_i^A=\delta_i^A\,Q(t)$, we must check whether the tensor perturbation to the gauge field, which is, by necessity, large when $x \sim 1$, does not produce a large backreaction on the background equations of motion. In order to estimate this backreaction, we will take advantage of our analytic solution for the tensor fluctuation to the gauge field, $t_R$, given in \eqref{tsol}. Of course this solution is merely a mode function, and we start by promoting $t_R$ and its conjugate $t_L$ to quantum operators. Starting with the definitions for $T_+,T_{\times}$, we arrive at the following forms for $\hat{t}_R$ and $\hat{t}_L$: 

\bea \label{ops t}
&& \hat{t}_R(z)= \int \frac{d^3k}{\left(2 \pi\right)^3}\left\{t_R(k)\,\hat{a}_R(k)\, e^{i \vec{k}.\vec{z}}+t_L^*(k)\, \hat{a}_L^{\dagger}(k)\, e^{-i \vec{k}.\vec{z}}\right\} \nonumber \\
&& \hat{t}_L(z)= \int \frac{d^3k}{\left(2 \pi\right)^3}\left\{t_L(k)\,\hat{a}_L(k)\, e^{i \vec{k}.\vec{z}}+t_R^*(k)\, \hat{a}_R^{\dagger}(k)\, e^{-i \vec{k}.\vec{z}}\right\}
\eea
where $\hat{t}_R=\hat{t}_L^*$ but $t_R \ne t_L^*$. The creation and annihilation operators satisfy:
\bea
&&\left[\hat{a}_R(p),\hat{a}_R^{\dagger}(q)\right]=\left(2 \pi\right)^3 \delta^3 \left(\vec{p}-\vec{q}\right) \nonumber \\
&&\left[\hat{a}_L(p),\hat{a}_L^{\dagger}(q)\right]=\left(2 \pi\right)^3 \delta^3 \left(\vec{p}-\vec{q}\right)
\eea
and all other combinations are zero. The mode function $t_R$ and its conjugate $t_R^*$ are given by \eqref{tsol} and $t_L,t_L^*$ are assumed to be negligible over the relevant region. The following integrals prove useful:
\bea
&& \langle 0 \lvert \hat{t}_R\, \hat{t}_L \rvert 0 \rangle = \int \frac{d^3k}{(2\pi)^3}\lvert t_R \rvert ^2 \nonumber \\
&& \langle 0 \lvert \hat{t}_L\, \hat{t}_R \rvert 0 \rangle = \int \frac{d^3k}{(2\pi)^3}\lvert t_L \rvert ^2 \sim 0 \nonumber \\
&& \langle 0 \lvert \hat{t}_R\, \partial_t\hat{t}_L \rvert 0 \rangle = \int \frac{d^3k}{(2\pi)^3}t_R\left(\partial_t t_R^*\right) \nonumber \\
&& \langle 0 \lvert \hat{t}_L\, \partial_t \hat{t}_R \rvert 0 \rangle = \int \frac{d^3k}{(2\pi)^3}t_L \left(\partial_t t_L^*\right) \sim 0 \nonumber \\
&& \langle 0 \lvert  \partial_t \hat{t}_L \, \hat{t}_R \rvert 0 \rangle = \int \frac{d^3k}{(2\pi)^3}\left(\partial_t t_L\right) t_L^* \sim 0 \nonumber \\
&& \langle 0 \lvert  \partial_t \hat{t}_R \, \hat{t}_L \rvert 0 \rangle = \int \frac{d^3k}{(2\pi)^3}\left(\partial_t t_R\right) t_R^* \nonumber \\
&& \langle 0 \lvert \hat{t}_R\, \partial_z \hat{t}_L \rvert 0 \rangle = \int \frac{d^3k}{(2\pi)^3}(-ik)\lvert t_R \rvert ^2 \nonumber \\
&& \langle 0 \lvert \hat{t}_L\, \partial_z \hat{t}_R \rvert 0 \rangle = \int \frac{d^3k}{(2\pi)^3}(-ik)\lvert t_L \rvert ^2 \sim 0 \nonumber \\
&& \langle 0 \lvert \partial_z\hat{t}_R\, \hat{t}_L \rvert 0 \rangle = \int \frac{d^3k}{(2\pi)^3}ik\lvert t_R \rvert ^2 \nonumber \\
&& \langle 0 \lvert \partial_z\hat{t}_L\, \hat{t}_R \rvert 0 \rangle = \int \frac{d^3k}{(2\pi)^3}ik \lvert t_L \rvert ^2 \sim 0 \nonumber \\
&& \langle 0 \lvert \partial_t \hat{t}_R\, \partial_t \hat{t}_L \rvert 0 \rangle = \int \frac{d^3k}{(2\pi)^3}\lvert \partial_t t_R \rvert ^2 \nonumber \\
&& \langle 0 \lvert \partial_t \hat{t}_L\, \partial_t \hat{t}_R \rvert 0 \rangle = \int \frac{d^3k}{(2\pi)^3}\lvert \partial_t t_L \rvert ^2 \sim 0 \nonumber \\
&& \langle 0 \lvert \partial_z \hat{t}_R\, \partial_z \hat{t}_L \rvert 0 \rangle = \int \frac{d^3k}{(2\pi)^3}k^2\lvert t_R \rvert ^2 \nonumber \\
&& \langle 0 \lvert \partial_z \hat{t}_L\, \partial_z \hat{t}_R \rvert 0 \rangle = \int \frac{d^3k}{(2\pi)^3}k^2\lvert t_L \rvert ^2 \sim 0 \nonumber \\
\eea
With these in hand, we can find equation of motion for $Q$, \eqref{GF1}, including the backreaction from $t_R$:
\bea\label{Q full}
&& \ddot{Q}+  3H   \dot{Q}  + Q\left(\dot{H}+2 H^2 \right)
+2{\tt g}^2Q^3-2{\tt g}\, Q^2\,H\,\xi_h
+2 H\,\xi_f  \left(  QH +  \dot{Q}  \right) \nonumber \\
&& +\frac{\g}{3\,a^2}\int \frac{d^3k}{(2 \pi)^3}\frac ka \lvert t_R \vert^2+\frac{\g\,\xi_h\,H}{3\, a^2} \int \frac{d^3k}{(2 \pi)^3} \lvert t_R \rvert^2 = 0
\eea
These additional terms are completely equivalent to the backreaction terms in \cite{DFF} and to make an estimate of their magnitude in terms of the effective mass of the gauge field, $\xi_Q$, we follow \cite{DFF} by defining:
\beq
{\cal T}_{BR}^Q\equiv \frac{\g\,\xi_h\,H}{3\, a^2} \int \frac{d^3k}{(2 \pi)^3} \lvert t_R \rvert^2+\frac{\g}{3\,a^2}\int \frac{d^3k}{(2 \pi)^3}\frac ka \lvert t_R \vert^2 \simeq \frac{\g\,H^3}{12 \pi^2} \left(\xi_h\,\beta_1(\xi_Q)+\beta_2(\xi_Q)\right)
\eeq
where 
\beq
\beta_1\left(\xi_Q\right)=\int^{x_{max}}_0 dx\, x \bigl \lvert i^{\beta} x^{\xi_f} W_{\beta,\alpha}\left(-2ix\right)  \bigr \rvert^2 \, ,
\eeq
\beq
\beta_2\left(\xi_Q\right)= \int^{x_{max}}_0 dx\, x^2 \bigl \lvert i^{\beta} x^{\xi_f} W_{\beta,\alpha}\left(-2ix\right)  \bigr \rvert^2 
\eeq
where we have used the analytic solution for $t_R$ given in \eqref{tsol}, and used the same cut-off described in \cite{DFF}, $x_{max}\equiv \xi_Q+\xi_h+\sqrt{\xi_Q^2+\xi_h^2}$ which encompasses the main region for which $t_R$ is enhanced by the transient instability near $x=1$. FIG \ref{backreaction} shows the evolution of the backreaction term, ${\cal T}_{BR}^Q$, plotted with the leading contributions to the equation of motion for $Q$. ${\cal T}^Q_{BR}$ is indeed small relative to the largest contribution given by $2 \g \, Q^2\, H\, \xi_h$. 

\begin{figure}[H]
	\centering
	\includegraphics[width=0.7\textwidth]{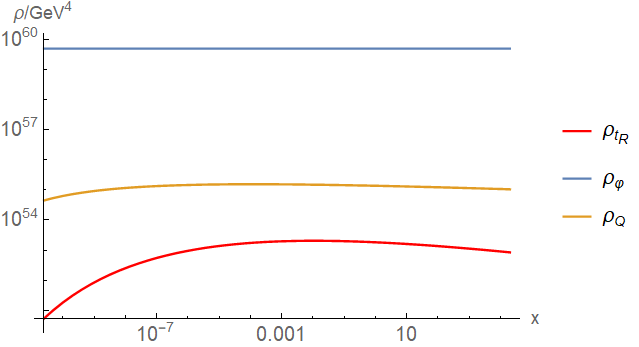}
	\caption{The evolution of the energy densities in $\phi$, $\rho_{\phi} \simeq V$ (blue); in $Q$, $\rho_Q = \frac 32 f \left[\left(H\,Q+\dot{Q}\right)^2+\g^2 Q^4\right]$ (orange); and in $t_R$, \eqref{rhotr} (red), for the example shown in section \ref{Sec3} plotted against $x=k/aH$.}
	\label{energy density br}
\end{figure}

We also want to check that the contribution to the energy density is low. The contribution to the energy density from the gauge tensor perturbation is found to be (exactly analogously to \cite{DFF})
\beq \label{rhotr}
\rho_{t_R}=\frac{1}{a^4} \int \frac{d^3k}{(2\pi)^3}\left\{\frac12 \bigl \lvert \partial_{\eta} t_R \bigr \rvert^2 +\left(\frac k2 + a\,\g\,Q\right)k\,\lvert t_R \rvert^2 \right\} \simeq f \frac{H^4}{8 \pi^2} {\cal I}_{t_R}
\eeq
where $\eta$ is conformal time and
\beq
{\cal I}_{t_R}=\int_0^{x_{max}}x^3 \left\{\bigl \lvert i^{\beta} \partial_x \left(x^{\xi_f} W_{\beta,\alpha}\left(-2 i x\right)\right) \bigr \rvert^2+\left(1+2 \frac{\xi_Q}{x}\right)\bigl \lvert i^{\beta} x^{\xi_f}  W_{\beta,\alpha} \left(-2 i x\right) \bigr \rvert^2 \right\} \, .
\eeq
Using this expression, in FIG. \ref{energy density br}, the evolution of $\rho_{t_R}$ is plotted alongside $\rho_{\phi} \simeq V$ and $\rho_Q = \frac 32 f \left[\left(H\,Q+\dot{Q}\right)^2+\g^2 Q^4\right]$. It is a sub-dominant contribution throughout. 

We have shown that the backreaction induced by the tensor mode $t_R$ on both the equation of motion for $Q$ as well as on the energy density is sub-dominant in the example shown in section \ref{Sec3}. Our original \"Ansatz for background form of $A_i^A$ is therefore consistent considering that the scalar perturbations to the gauge field are small.

\section{Discussion}\label{disc}

Spectator Chromonatural Inflation \cite{DP,Adshead:2012kp,Adshead:2012qe,DFF,Fujita17} involving non-Abelian gauge field backgrounds offers an appealing mechanism to produce a gravitational wave spectrum with a preferred handedness that can be produced at observable levels even when no field makes a super-Planckian field excursion, thus evading the Lyth bound.

As we discussed in this paper, designing field theory models with the correct  properties  to realise   satisfactory (S)CNI seems rather challenging \cite{Reece17,Reece}. 
Supergravity and string theory models  of inflation offer a natural framework where one can study possible embeddings of SCNI that may overcome some of these challenges. 
A first look at the supergravity embedding of CNI was taken in \cite{DAgata}, while 
in string theory, a proposal to embed SCNI was first made in \cite{McDA}.

In this paper we took a further step towards a successful embedding of SCNI in string theory.  We first introduced in section \ref{Sec2} a generalised multifield model that incorporates the ingredients that arise in supergravity and string theory models with scalars and gauge fields. Namely a non-trivial scalar metric, potential coupling of the scalars to the gauge field besides the axion. We  discussed this more general set-up in comparison with the SCNI and discussed the required constraints to achieve a successful inflationary evolution. 

We then used our results in section \ref{Sec2} to  analyse the careful embedding of SCNI within a string theory model of inflation in section \ref{Sec3}. 
As in \cite{McDA} we used K\"ahler inflation as a host model, which naturally predicts a low value for the tensor-to-scalar ratio $r\lesssim 10^{-7}$ \cite{KI1,KI2}, well outside observable limits. As we discussed, K\"ahler inflation in its original form, cannot be used to realise SCNI. Thus, similarly to \cite{McDA}, we introduced as a spectator sector magnetised D7-branes, multiply wrapping four cycles, parameterised by a K\"ahler modulus $T_4$ and a $C_2$ axion,  $b$, coupled to the non-Abelian gauge fields living on the magnetised multiply wrapped D7-branes. 
The parameters  available in this set-up are the magnetic field on the spectator D7-branes, $m$, its wrapping number, $n$, and the degree of the gauge group, $N$. 
From the cosmological point of view, we were interested in realising successfully  three specific goals:  successful inflation, a sustained gauge field to successfully source gravitational waves at the observable level, $r\sim 10^{-3}$, and a controllable backreaction from the tensorial gauge fluctuations. We therefore fixed the parameters $(m,N,n)$ in order to realise these goals. 

Once we fixed the parameters, we first checked the moduli stabilisation of our four K\"ahler moduli system:  that is, three moduli necessary for K\"ahler inflation plus the fourth modulus required as a spectator. We  fixed all moduli except the inflaton, $\tau_2$, and the spectators, $\tau_4$, $b$ to their minima\footnote{We checked that this can be done consistently as in the original K\"ahler Inflation model \cite{KI1}.}.
We were then left with a general multifield system described from the field theory point of view by the action \eqref{action2}, where the field space metric $\gamma_{ab}$ is given by\footnote{Where we  are working in the large volume limit.} \eqref{gamma},  the scalar potential  is given by \eqref{InfPot},  the couplings $f(\phi^a), h(\phi^a)$ are given by $f = \tau_4$, $h= m \,b$,   the gauge field $F=dA- {\tt g} A\wedge A$ and  we defined the ``effective gauge coupling" as $  {\tt g} = 1/\sqrt{n N/2}$

As we discussed in sections \ref{Sec2} and \ref{Sec3}, a successful period of inflation and large enough enhancement of the gravitational wave spectrum impose two conditions on the parameters, which can be fixed by two of the three parameters  in the model, namely $m$ and $N$. However, keeping the backreaction under control imposes a third condition, which then fixes the third parameter, $n$.  

We  evolve the full four-field system, $(\tau_2, \tau_4, b, Q)$,  numerically and showed our results in sections \ref{Sec3} and \ref{Sec4}.  We show  in figure \ref{Estima_r}  the enhancement of $r$ for different values of $ {\tt g}$ as a function of the parameter $\xi_Q$, while in figure \ref{BackR}, we see  the backreaction estimate for three values of $ {\tt g}$ as a function of $\xi_Q$. As we have discussed, for a fixed value of the effective coupling, the larger $\xi_Q$, the larger the enhancement, but also the backreaction (see also figure \ref{backreaction}).  We  show the dependence of $ {\tt g} $ as a function of $N$ in figure \ref{gofN}. 

As we discussed in section \ref{Sec3}, since the spectator axion is not canonically normalised, and its kinetic term depends on $\tau_4$, a dynamical field, we cannot define a decay ``constant" in the usual way. However, we can define an ``instantaneous decay constant", as in \cite{FatI}, which we show in figure \ref{epsvBKI2} in Planck units and turns out to be sub-Planckian as expected, $f_c\sim 10^{-3}\Mp$. On the other hand, because the host inflationary model is K\"ahler inflation, the field excursion is sub-Planckian as well, $\Delta\phi \sim 0.2 \Mp$ and the Lyth bound is evaded, since the enhanced tensor spectrum gives $r = 2.29\times 10^{-3}$ (see section \ref{Sec4}). 
The actual gauge kinetic coupling is field dependent and set by $\tau_4$. At the minimum of $\tau_4$, this is given by $g^2=1/\langle\tau_4\rangle\sim \frac1{15}$, and remember that it is displaced only a small distance from its minimum, so an instantaneous gauge coupling defined at time $t$  differs little  from its value at the minimum. 

\begin{figure}[H]
	\centering
	\includegraphics[width=0.7\textwidth]{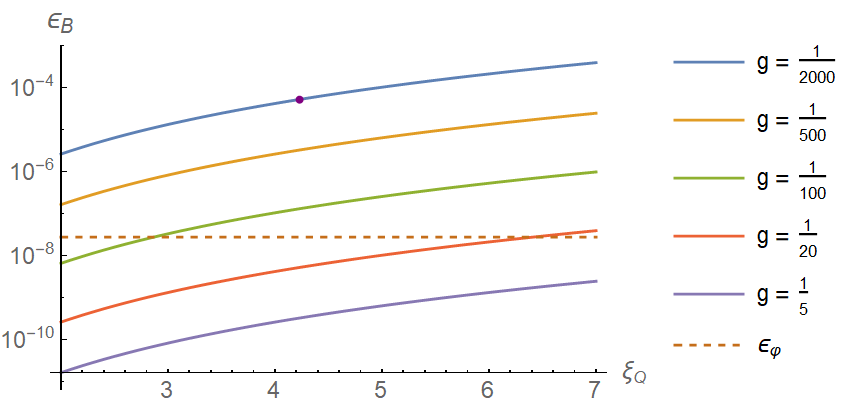}
	\caption{The value of $\epsilon_B = f \frac{\g^2Q^4}{\Mp^2 H^2} =f \frac{H^2 \xi_Q^4}{\g^2 \Mp^2}$ plotted against $\xi_Q= \frac{\g Q}{H}$. If one wishes to achieve the same value of $\xi_Q$ after reducing $\g$ to mitigate the backreaction, $\epsilon_B$ will be larger. The dashed line shows the value of $\epsilon_{\varphi}$ $60$ e-folds before the end of inflation for the example shown in this paper. Similarly the dot corresponds to the values for $\xi_Q$ and $\epsilon_B$ in the example shown in this paper $60$ e-folds before the end of inflation.}
	\label{epsBvxiQ}
\end{figure}

\begin{figure}[H]
	\centering
	\includegraphics[width=0.4\textwidth]{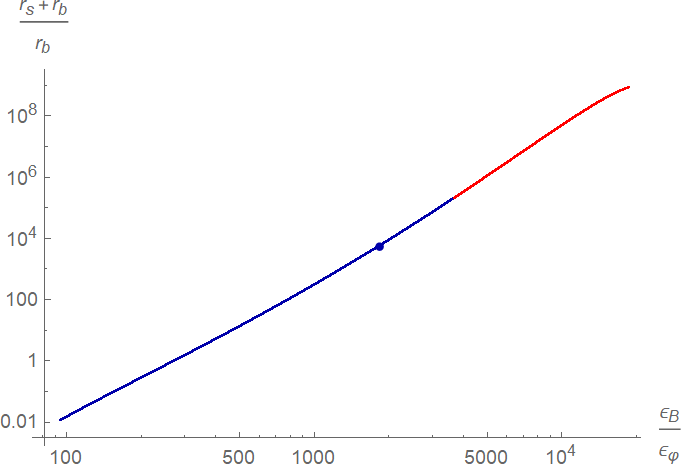}
	\includegraphics[width=0.4\textwidth]{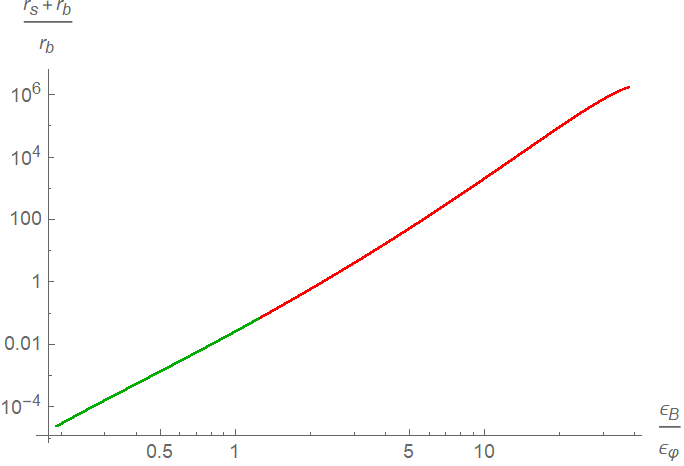}
	\caption{Two plots showing how the enhancement factor, $\frac{r_s+r_{b}}{r_b}$, varies relative to the ratio of $\frac{\epsilon_B}{\epsilon_{\varphi}}$ where  $\epsilon_{\varphi} = 2.80 \times 10^{-8}$, the value it takes $60$ e-folds before the end of inflation. The left plot corresponds to $\g = \frac 1{2000}$, the value of $\g$ used in the example of this paper and the right plot shows $\g = \frac1{100}$. The dot corresponds to the values in the example shown in this paper. The red part of the lines corresponds to parameter space where the backreaction is too large. It can be seen immediately that with $\g = \frac1{100}$, it is impossible to get a substantial enhancement without incurring excessive backreaction. Because of the low value of $\epsilon_{\varphi}$ in K\"ahler inflation, in order to get a large enhancement with controlled backreaction, we require $\g$ small, which leads to a large value of $\epsilon_B$ relative to $\epsilon_{\varphi}$.}
	\label{epsbp}
\end{figure}

An important feature of the model from the cosmological point of view, is that it realises a very mild version of the proposal in \cite{Fujita17} to enhance the gravitational wave spectrum. Indeed, the model requires a specific  hierarchy  in the slow-roll parameters, namely  $\epsilon\sim \epsilon_B\gg \epsilon_\varphi$ in order to ensure such a large enhancement. To see why this is we plot in FIG. \ref{epsBvxiQ} how $\epsilon_B$ is affected by reducing the value of $\g$. Despite the fact that, naively, it seems $\epsilon_B  = f \frac{\g^2Q^4}{\Mp^2 H^2} \propto \g^2$, in fact if we plot it against $\xi_Q$, we see that $\epsilon_B =f \frac{H^2 \xi_Q^4}{\g^2 \Mp^2}$. This fact has the important consequence that if we wish to have a model that produces a relatively large value of $\xi_Q$ (e.g. in our case $\xi_Q \sim 4$) so that we can get a large enough enhancement to the tensor-to-scalar (see FIG. \ref{Estima_r}) to achieve $r \sim 10^{-3}$, whilst having a small value of $\g$ so as not to produce too large a backreaction (see FIG. \ref{BackR}), it is unavoidable that $\epsilon \sim \epsilon_B > \epsilon_{\varphi}$. This point is further emphasised in FIG. \ref{epsbp} where we plot the enhancement factor against the ratio of $\epsilon_B$ to $\epsilon_{\varphi}$ for two values of $\g$. In \cite{DFF}, their inflationary model already predicts a background tensor-to-scalar ratio of $r_{b} \sim 10^{-3}$; they have a much larger value of $\epsilon_{\varphi}$ than in our example; and their model predicts a considerably smaller enhancement factor $\frac{r_s+r_{b}}{r_{b}} \sim 20$ (compared to $\frac{r_s+r_{b}}{r_{b}} \sim 5000$ in our example). To achieve this enhancement, they only require $\xi_Q \sim 3.4$ compared to our $\xi_Q \sim 4.2$ and as discussed the backreaction scales exponentially in $\xi_Q$. For these reasons, in \cite{DFF}, they are able to choose a relatively large value for $\g = 1.11 \times 10^{-2}$ compared to our $\g = \frac 1{2000}$, and therefore they are able to satisfy $\epsilon \sim \epsilon_{\varphi} \gg \epsilon_B$. In this regard our example has more in common with \cite{Fujita17} where they demonstrate that one can achieve exceptionally large enhancements whilst controlling the backreaction if $\g$ is taken to be small enough and it is allowed that $\epsilon \sim \epsilon_B$. In the specific example in  \cite{Fujita17}, they achieve an enhancement of $\frac{r_s+r_{b}}{r_{b}} \sim 10^{68}$ and have $\epsilon \sim \epsilon_B \sim 10^{-2}$.  

The fact that we have $\epsilon \sim \epsilon_B$ may be seen as a problem. However as in \cite{Fujita17}, we assume that scalar power spectrum, $\mathcal{P}_S$, is dominated by the scalar field perturbations and the gauge field perturbations therefore contribute negligibly to $\mathcal{P}_S$. This in turn allows us to assume that the scalar power spectrum is well approximated by $\mathcal{P}_S = \frac{H^2}{8 \pi^2 \epsilon_{\varphi}}$ and consequently that $n_s = 1 - 2 \epsilon - \eta_{\varphi}$. This of course ensures that the important inflationary predictions of K\"ahler inflation are not spoiled. In figures \ref{scalar perts}-\ref{scalar perts long}, we show that the scalar field perturbations are much larger in magnitude than the scalar perturbations of the gauge field.


Comparing to the phenomenological models in the literature, we have seen that a relatively simple string theory construction has enough parameters to account for the cosmological constraints required to realised the our goals. At least one of these parameters can be easily incorporated in field theory models, namely rather than working with $SU(2)$, one can work with $SU(N)$, introducing an effective coupling that can be made small, at the price of a large degree. However it is not clear whether there are field theory equivalents of the magnetic flux and the D7-brane wrapping number.

Let us now discuss in more detail the values of the model parameters $(m,N,n)$. As we saw, for a successful evolution we require a large magnetic flux $m= 10^4$. Requiring that the gauge field is sustained for enough e-folds to enhance the tensor spectrum then requires that the instantaneous decay constant $f_{c}\sim 10^{-3}\Mp$, which fixes $N\sim3 \times 10^5$. Finally control on the backreaction requires a specific value for the effective decay constant $ {\tt g}\sim \frac1{2000}$, which fixes $n= 25$. 
The winding number is of a similar order to the values    used in e.g.~\cite{Long}. The magnetic flux on the other hand is much larger, and may backreact on the geometry. Moreover,   the gauge group degree required  is very large $N\sim 10^{5}$. Since $N$ is basically the number of D7-branes, such a large number of them  may again backreact on the full geometry. In any case, it is not clear that such a large number for $N$ can be realised in any realistic large volume compactification. So we see this as the biggest challenge of the construction. Remember however, that we have required three very specific cosmological objectives and the large value of $N$ is needed to realise these. 

A final comment on possible constraints from the weak gravity conjecture  (WGC) \cite{WGC} in the $SU(N)$ case \cite{Reece17}. 
As we mentioned, the gauge coupling is set by $g^2\sim1/\tau_4$, and given the large value of $N$, the cutoff implied by the WGC for $SU(N)$ is of order $\Lambda_{QG}\sim g \Mp\sim 0.3\Mp$. As we have also mentioned, we cannot define a decay constant similarly to the model discussed in \cite{FatI}, so it is not clear how the axionic version of the WGC would apply. In any case, as we have shown, we can define an instantaneous decay constant, which is sub-Planckian, as it is the field excursion. 

Let us finally comment on using a different string inflation model as a host, namely Fibre inflation \cite{FI1}. Fibre inflation  already predicts a relatively large  value  for the tensor-to-scalar ratio from the vacuum fluctuations, namely $r_b\sim 10^{-3}$. In this case, a similar construction to the one described in the present work for the spectator sector can be used  to enhance the inflationary tensor mode to about $r\sim 0.01$ with the parameter choice $(m,N,n)=(500,5000,1) $ \cite{Thesis} and therefore its chirality may be accessible to observations \cite{Gluscevic:2010vv}. Since the enhancement is much smaller, the required values for the parameters are also reduced, being 20 times smaller. However, the gauge group degree  remains still rather large. 
 
In summary, we find our results very interesting in that string theory models of inflation contain potentially all the necessary ingredients to realise SCNI. However, it is clear that requiring an appealing cosmology  imposes strong constraints on the parameters, which may not be realisable in realistic constructions. On the other hand, the construction is rich enough and has other cosmological implications at different scales than the CMB, which may relax the constraints on the parameters. We leave for future work further investigations of this aspect.

\section*{Acknowledgements}
We are grateful to Michele Cicoli, Emanuela Dimastrogiovanni, Matteo Fasiello, Elisa Ferreira, Anshuman Maharana,  Evan McDonough, Ryo Namba, Carlos N\'u\~nez,  Susha Parameswaran and Fernando Quevedo for discussions and Ogan \"Ozsoy for collaboration at early stages of this work. 
J. H. was supported by a Swansea College of Science Doctoral Training Centre (DTC)  Research  Scholarship  funded  partly  by  Swansea University and partly by the STFC.  IZ  and GT are partially supported by STFC, grant ST/P00055X/1.

\addcontentsline{toc}{section}{References}
\bibliographystyle{utphys}

\bibliography{refs}

\providecommand{\href}[2]{#2}\begingroup\raggedright\begin{thebibliography}{10}

\bibitem{Planck18}
{\bf Planck} Collaboration, Y.~Akrami {\em et al.}, ``{Planck 2018 results. X.
  Constraints on inflation}'', \href{http://arxiv.org/abs/1807.06211}{{\tt
  arXiv:1807.06211 [astro-ph.CO]}}.

\bibitem{ZU}
M.~Zaldarriaga and U.~Seljak, ``{An all sky analysis of polarization in the
  microwave background}'',
  \href{http://dx.doi.org/10.1103/PhysRevD.55.1830}{{\em Phys. Rev. D} {\bf 55}
  (1997)  1830--1840}, \href{http://arxiv.org/abs/astro-ph/9609170}{{\tt
  arXiv:astro-ph/9609170}}.

\bibitem{KamionB}
M.~Kamionkowski, A.~Kosowsky, and A.~Stebbins, ``{Statistics of cosmic
  microwave background polarization}'',
  \href{http://dx.doi.org/10.1103/PhysRevD.55.7368}{{\em Phys. Rev. D} {\bf 55}
  (1997)  7368--7388}, \href{http://arxiv.org/abs/astro-ph/9611125}{{\tt
  arXiv:astro-ph/9611125}}.

\bibitem{SPT3G}
{\bf SPT-3G} Collaboration, B.~Benson {\em et al.}, ``{SPT-3G: A
  Next-Generation Cosmic Microwave Background Polarization Experiment on the
  South Pole Telescope}'', \href{http://dx.doi.org/10.1117/12.2057305}{{\em
  Proc. SPIE Int. Soc. Opt. Eng.} {\bf 9153} (2014)  91531P},
  \href{http://arxiv.org/abs/1407.2973}{{\tt arXiv:1407.2973 [astro-ph.IM]}}.

\bibitem{simons}
{\bf Simons Observatory} Collaboration, J.~Aguirre {\em et al.}, ``{The Simons
  Observatory: Science goals and forecasts}'',
  \href{http://dx.doi.org/10.1088/1475-7516/2019/02/056}{{\em JCAP} {\bf 1902}
  (2019)  056},
\href{http://arxiv.org/abs/1808.07445}{{\tt arXiv:1808.07445 [astro-ph.CO]}}.

\bibitem{cmb-s4}
{\bf CMB-S4} Collaboration, K.~N. Abazajian {\em et al.}, ``{CMB-S4 Science
  Book, First Edition}'',
\href{http://arxiv.org/abs/1610.02743}{{\tt arXiv:1610.02743 [astro-ph.CO]}}.

\bibitem{class}
T.~Essinger-Hileman {\em et al.}, ``{CLASS: The Cosmology Large Angular Scale
  Surveyor}'', \href{http://dx.doi.org/10.1117/12.2056701}{{\em Proc. SPIE Int.
  Soc. Opt. Eng.} {\bf 9153} (2014)  91531I},
\href{http://arxiv.org/abs/1408.4788}{{\tt arXiv:1408.4788 [astro-ph.IM]}}.

\bibitem{Lbird}
T.~Matsumura {\em et al.}, ``{Mission design of LiteBIRD}'',
  \href{http://arxiv.org/abs/1311.2847}{{\tt arXiv:1311.2847 [astro-ph.IM]}}.
[J. Low. Temp. Phys.176,733(2014)].

\bibitem{pico}
{\bf NASA PICO} Collaboration, S.~Hanany {\em et al.}, ``{PICO: Probe of
  Inflation and Cosmic Origins}'',
\href{http://arxiv.org/abs/1902.10541}{{\tt arXiv:1902.10541 [astro-ph.IM]}}.

\bibitem{Lyth}
D.~H. Lyth, ``{What would we learn by detecting a gravitational wave signal in
  the cosmic microwave background anisotropy?}'',
  \href{http://dx.doi.org/10.1103/PhysRevLett.78.1861}{{\em Phys. Rev. Lett.}
  {\bf 78} (1997)  1861--1863}, \href{http://arxiv.org/abs/hep-ph/9606387}{{\tt
  arXiv:hep-ph/9606387}}.

\bibitem{Lotfi}
L.~Boubekeur and D.~H. Lyth, ``{Hilltop inflation}'',
  \href{http://dx.doi.org/10.1088/1475-7516/2005/07/010}{{\em JCAP} {\bf 07}
  (2005)  010}, \href{http://arxiv.org/abs/hep-ph/0502047}{{\tt
  arXiv:hep-ph/0502047}}.

\bibitem{Garcia-Bellido:2014wfa}
J.~Garcia-Bellido, D.~Roest, M.~Scalisi, and I.~Zavala, ``{Lyth bound of
  inflation with a tilt}'',
  \href{http://dx.doi.org/10.1103/PhysRevD.90.123539}{{\em Phys. Rev. D} {\bf
  90} (2014) no.~12, 123539}, \href{http://arxiv.org/abs/1408.6839}{{\tt
  arXiv:1408.6839 [hep-th]}}.

\bibitem{Gong}
J.-O. Gong, ``{Multi-field inflation and cosmological perturbations}'',
  \href{http://dx.doi.org/10.1142/S021827181740003X}{{\em Int. J. Mod. Phys. D}
  {\bf 26} (2016) no.~01, 1740003}, \href{http://arxiv.org/abs/1606.06971}{{\tt
  arXiv:1606.06971 [gr-qc]}}.

\bibitem{MSJS}
A.~Maleknejad, M.~M. Sheikh-Jabbari, and J.~Soda, ``{Gauge Fields and
  Inflation}'', \href{http://dx.doi.org/10.1016/j.physrep.2013.03.003}{{\em
  Phys. Rept.} {\bf 528} (2013)  161--261},
\href{http://arxiv.org/abs/1212.2921}{{\tt arXiv:1212.2921 [hep-th]}}.

\bibitem{MSJ1}
A.~Maleknejad and M.~Sheikh-Jabbari, ``{Gauge-flation: Inflation From
  Non-Abelian Gauge Fields}'',
  \href{http://dx.doi.org/10.1016/j.physletb.2013.05.001}{{\em Phys. Lett. B}
  {\bf 723} (2013)  224--228}, \href{http://arxiv.org/abs/1102.1513}{{\tt
  arXiv:1102.1513 [hep-ph]}}.

\bibitem{MSJ2}
A.~Maleknejad and M.~Sheikh-Jabbari, ``{Non-Abelian Gauge Field Inflation}'',
  \href{http://dx.doi.org/10.1103/PhysRevD.84.043515}{{\em Phys. Rev. D} {\bf
  84} (2011)  043515}, \href{http://arxiv.org/abs/1102.1932}{{\tt
  arXiv:1102.1932 [hep-ph]}}.

\bibitem{SJ}
M.~Sheikh-Jabbari, ``{Gauge-flation Vs Chromo-Natural Inflation}'',
  \href{http://dx.doi.org/10.1016/j.physletb.2012.09.014}{{\em Phys. Lett. B}
  {\bf 717} (2012)  6--9}, \href{http://arxiv.org/abs/1203.2265}{{\tt
  arXiv:1203.2265 [hep-th]}}.

\bibitem{Adshead:2012kp}
P.~Adshead and M.~Wyman, ``{Chromo-Natural Inflation: Natural inflation on a
  steep potential with classical non-Abelian gauge fields}'',
  \href{http://dx.doi.org/10.1103/PhysRevLett.108.261302}{{\em Phys.\ Rev.\
  Lett.} {\bf 108} (2012)  261302}, \href{http://arxiv.org/abs/1202.2366}{{\tt
  arXiv:1202.2366 [hep-th]}}.

\bibitem{Adshead:2012qe}
P.~Adshead and M.~Wyman, ``{Gauge-flation trajectories in Chromo-Natural
  Inflation}'', \href{http://dx.doi.org/10.1103/PhysRevD.86.043530}{{\em Phys.
  Rev.} {\bf D86} (2012)  043530},
\href{http://arxiv.org/abs/1203.2264}{{\tt arXiv:1203.2264 [hep-th]}}.

\bibitem{MNSJ}
A.~Maleknejad, M.~Noorbala, and M.~Sheikh-Jabbari, ``{Leptogenesis in
  inflationary models with non-Abelian gauge fields}'',
  \href{http://dx.doi.org/10.1007/s10714-018-2435-8}{{\em Gen. Rel. Grav.} {\bf
  50} (2018) no.~9, 110}, \href{http://arxiv.org/abs/1208.2807}{{\tt
  arXiv:1208.2807 [hep-th]}}.

\bibitem{Biagetti:2013kwa}
M.~Biagetti, M.~Fasiello, and A.~Riotto, ``{Enhancing Inflationary Tensor Modes
  through Spectator Fields}'',
  \href{http://dx.doi.org/10.1103/PhysRevD.88.103518}{{\em Phys. Rev. D} {\bf
  88} (2013)  103518}, \href{http://arxiv.org/abs/1305.7241}{{\tt
  arXiv:1305.7241 [astro-ph.CO]}}.

\bibitem{Biagetti:2014asa}
M.~Biagetti, E.~Dimastrogiovanni, M.~Fasiello, and M.~Peloso, ``{Gravitational
  Waves and Scalar Perturbations from Spectator Fields}'',
  \href{http://dx.doi.org/10.1088/1475-7516/2015/04/011}{{\em JCAP} {\bf 04}
  (2015)  011}, \href{http://arxiv.org/abs/1411.3029}{{\tt arXiv:1411.3029
  [astro-ph.CO]}}.

\bibitem{Fujita:2014oba}
T.~Fujita, J.~Yokoyama, and S.~Yokoyama, ``{Can a spectator scalar field
  enhance inflationary tensor mode?}'',
  \href{http://dx.doi.org/10.1093/ptep/ptv037}{{\em PTEP} {\bf 2015} (2015)
  043E01}, \href{http://arxiv.org/abs/1411.3658}{{\tt arXiv:1411.3658
  [astro-ph.CO]}}.

\bibitem{Bartolo:2016ami}
N.~Bartolo {\em et al.}, ``{Science with the space-based interferometer LISA.
  IV: Probing inflation with gravitational waves}'',
  \href{http://dx.doi.org/10.1088/1475-7516/2016/12/026}{{\em JCAP} {\bf 1612}
  (2016)  026},
\href{http://arxiv.org/abs/1610.06481}{{\tt arXiv:1610.06481 [astro-ph.CO]}}.

\bibitem{Namba:2015gja}
R.~Namba, M.~Peloso, M.~Shiraishi, L.~Sorbo, and C.~Unal, ``{Scale-dependent
  gravitational waves from a rolling axion}'',
  \href{http://dx.doi.org/10.1088/1475-7516/2016/01/041}{{\em JCAP} {\bf 01}
  (2016)  041}, \href{http://arxiv.org/abs/1509.07521}{{\tt arXiv:1509.07521
  [astro-ph.CO]}}.

\bibitem{Peloso:2016gqs}
M.~Peloso, L.~Sorbo, and C.~Unal, ``{Rolling axions during inflation:
  perturbativity and signatures}'',
  \href{http://dx.doi.org/10.1088/1475-7516/2016/09/001}{{\em JCAP} {\bf 09}
  (2016)  001}, \href{http://arxiv.org/abs/1606.00459}{{\tt arXiv:1606.00459
  [astro-ph.CO]}}.

\bibitem{ogan}
O.~Ozsoy, ``{Gravitational Waves from a Rolling Axion Monodromy}'',
  \href{http://arxiv.org/abs/2005.10280}{{\tt arXiv:2005.10280 [astro-ph.CO]}}.

\bibitem{Adshead:2013qp}
P.~Adshead, E.~Martinec, and M.~Wyman, ``{Gauge fields and inflation: Chiral
  gravitational waves, fluctuations, and the Lyth bound}'',
  \href{http://dx.doi.org/10.1103/PhysRevD.88.021302}{{\em Phys. Rev. D} {\bf
  88} (2013) no.~2, 021302}, \href{http://arxiv.org/abs/1301.2598}{{\tt
  arXiv:1301.2598 [hep-th]}}.

\bibitem{Adshead:2013nka}
P.~Adshead, E.~Martinec, and M.~Wyman, ``{Perturbations in Chromo-Natural
  Inflation}'', \href{http://dx.doi.org/10.1007/JHEP09(2013)087}{{\em JHEP}
  {\bf 09} (2013)  087}, \href{http://arxiv.org/abs/1305.2930}{{\tt
  arXiv:1305.2930 [hep-th]}}.

\bibitem{NI}
K.~Freese, J.~A. Frieman, and A.~V. Olinto, ``{Natural inflation with pseudo -
  Nambu-Goldstone bosons}'',
  \href{http://dx.doi.org/10.1103/PhysRevLett.65.3233}{{\em Phys. Rev. Lett.}
  {\bf 65} (1990)  3233--3236}.

\bibitem{OS}
I.~Obata and J.~Soda, ``{Chiral primordial gravitational waves from dilaton
  induced delayed chromonatural inflation}'',
  \href{http://dx.doi.org/10.1103/PhysRevD.95.109903,
  10.1103/PhysRevD.93.123502}{{\em Phys. Rev.} {\bf D93} (2016) no.~12,
  123502}, \href{http://arxiv.org/abs/1602.06024}{{\tt arXiv:1602.06024
  [hep-th]}}.
[Addendum: Phys. Rev.D95,no.10,109903(2017)].

\bibitem{DFF}
E.~Dimastrogiovanni, M.~Fasiello, and T.~Fujita, ``{Primordial Gravitational
  Waves from Axion-Gauge Fields Dynamics}'',
  \href{http://dx.doi.org/10.1088/1475-7516/2017/01/019}{{\em JCAP} {\bf 1701}
  (2017) no.~01, 019},
\href{http://arxiv.org/abs/1608.04216}{{\tt arXiv:1608.04216 [astro-ph.CO]}}.

\bibitem{Azadeh}
A.~Maleknejad, ``{Axion Inflation with an SU(2) Gauge Field: Detectable Chiral
  Gravity Waves}'', \href{http://dx.doi.org/10.1007/JHEP07(2016)104}{{\em JHEP}
  {\bf 07} (2016)  104},
\href{http://arxiv.org/abs/1604.03327}{{\tt arXiv:1604.03327 [hep-ph]}}.

\bibitem{Reece17}
B.~Heidenreich, M.~Reece, and T.~Rudelius, ``{The Weak Gravity Conjecture and
  Emergence from an Ultraviolet Cutoff}'',
  \href{http://dx.doi.org/10.1140/epjc/s10052-018-5811-3}{{\em Eur. Phys. J. C}
  {\bf 78} (2018) no.~4, 337}, \href{http://arxiv.org/abs/1712.01868}{{\tt
  arXiv:1712.01868 [hep-th]}}.

\bibitem{Reece}
P.~Agrawal, J.~Fan, and M.~Reece, ``{Clockwork Axions in Cosmology: Is
  Chromonatural Inflation Chrononatural?}'',
  \href{http://dx.doi.org/10.1007/JHEP10(2018)193}{{\em JHEP} {\bf 10} (2018)
  193},
\href{http://arxiv.org/abs/1806.09621}{{\tt arXiv:1806.09621 [hep-th]}}.

\bibitem{Fujita17}
T.~Fujita, R.~Namba, and Y.~Tada, ``{Does the detection of primordial
  gravitational waves exclude low energy inflation?}'',
  \href{http://dx.doi.org/10.1016/j.physletb.2017.12.014}{{\em Phys. Lett.}
  {\bf B778} (2018)  17--21},
\href{http://arxiv.org/abs/1705.01533}{{\tt arXiv:1705.01533 [astro-ph.CO]}}.

\bibitem{MK}
A.~Maleknejad and E.~Komatsu, ``{Production and Backreaction of Spin-2
  Particles of $SU(2)$ Gauge Field during Inflation}'',
  \href{http://dx.doi.org/10.1007/JHEP05(2019)174}{{\em JHEP} {\bf 05} (2019)
  174},
\href{http://arxiv.org/abs/1808.09076}{{\tt arXiv:1808.09076 [hep-ph]}}.

\bibitem{DAgata}
G.~Dall'Agata, ``{Chromo-Natural inflation in Supergravity}'',
  \href{http://dx.doi.org/10.1016/j.physletb.2018.05.020}{{\em Phys. Lett.}
  {\bf B782} (2018)  139--142},
\href{http://arxiv.org/abs/1804.03104}{{\tt arXiv:1804.03104 [hep-th]}}.

\bibitem{McDA}
E.~McDonough and S.~Alexander, ``{Observable Chiral Gravitational Waves from
  Inflation in String Theory}'',
  \href{http://dx.doi.org/10.1088/1475-7516/2018/11/030}{{\em JCAP} {\bf 1811}
  (2018) no.~11, 030},
\href{http://arxiv.org/abs/1806.05684}{{\tt arXiv:1806.05684 [hep-th]}}.

\bibitem{Long}
C.~Long, L.~McAllister, and P.~McGuirk, ``{Aligned Natural Inflation in String
  Theory}'', \href{http://dx.doi.org/10.1103/PhysRevD.90.023501}{{\em Phys.
  Rev. D} {\bf 90} (2014)  023501}, \href{http://arxiv.org/abs/1404.7852}{{\tt
  arXiv:1404.7852 [hep-th]}}.

\bibitem{Ido}
I.~Ben-Dayan, F.~G. Pedro, and A.~Westphal, ``{Towards Natural Inflation in
  String Theory}'', \href{http://dx.doi.org/10.1103/PhysRevD.92.023515}{{\em
  Phys. Rev. D} {\bf 92} (2015) no.~2, 023515},
  \href{http://arxiv.org/abs/1407.2562}{{\tt arXiv:1407.2562 [hep-th]}}.

\bibitem{LV1}
V.~Balasubramanian, P.~Berglund, J.~P. Conlon, and F.~Quevedo, ``{Systematics
  of moduli stabilisation in Calabi-Yau flux compactifications}'',
  \href{http://dx.doi.org/10.1088/1126-6708/2005/03/007}{{\em JHEP} {\bf 03}
  (2005)  007},
\href{http://arxiv.org/abs/hep-th/0502058}{{\tt arXiv:hep-th/0502058
  [hep-th]}}.

\bibitem{LV2}
J.~P. Conlon, F.~Quevedo, and K.~Suruliz, ``{Large-volume flux
  compactifications: Moduli spectrum and D3/D7 soft supersymmetry breaking}'',
  \href{http://dx.doi.org/10.1088/1126-6708/2005/08/007}{{\em JHEP} {\bf 08}
  (2005)  007},
\href{http://arxiv.org/abs/hep-th/0505076}{{\tt arXiv:hep-th/0505076
  [hep-th]}}.

\bibitem{KI1}
J.~P. Conlon and F.~Quevedo, ``{Kahler moduli inflation}'',
  \href{http://dx.doi.org/10.1088/1126-6708/2006/01/146}{{\em JHEP} {\bf 01}
  (2006)  146},
\href{http://arxiv.org/abs/hep-th/0509012}{{\tt arXiv:hep-th/0509012
  [hep-th]}}.

\bibitem{KI2}
J.~J. Blanco-Pillado, D.~Buck, E.~J. Copeland, M.~Gomez-Reino, and N.~J. Nunes,
  ``{Kahler Moduli Inflation Revisited}'',
  \href{http://dx.doi.org/10.1007/JHEP01(2010)081}{{\em JHEP} {\bf 01} (2010)
  081},
\href{http://arxiv.org/abs/0906.3711}{{\tt arXiv:0906.3711 [hep-th]}}.

\bibitem{Caldwell:2017chz}
R.~R. Caldwell and C.~Devulder, ``{Axion Gauge Field Inflation and
  Gravitational Leptogenesis: A Lower Bound on B Modes from the
  Matter-Antimatter Asymmetry of the Universe}'',
  \href{http://dx.doi.org/10.1103/PhysRevD.97.023532}{{\em Phys. Rev.} {\bf
  D97} (2018) no.~2, 023532},
\href{http://arxiv.org/abs/1706.03765}{{\tt arXiv:1706.03765 [astro-ph.CO]}}.

\bibitem{Caldwell:2018feo}
R.~R. Caldwell and C.~Devulder, ``{Gravitational Wave Opacity from Gauge Field
  Dark Energy}'',
\href{http://arxiv.org/abs/1802.07371}{{\tt arXiv:1802.07371 [gr-qc]}}.

\bibitem{Achucarro:2015rfa}
A.~Achucarro, V.~Atal, and Y.~Welling, ``{On the viability of $m^2\phi^2$ and
  natural inflation}'',
  \href{http://dx.doi.org/10.1088/1475-7516/2015/07/008}{{\em JCAP} {\bf 07}
  (2015)  008}, \href{http://arxiv.org/abs/1503.07486}{{\tt arXiv:1503.07486
  [astro-ph.CO]}}.

\bibitem{Brown:2017osf}
A.~R. Brown, ``{Hyperbolic Inflation}'',
  \href{http://dx.doi.org/10.1103/PhysRevLett.121.251601}{{\em Phys. Rev.
  Lett.} {\bf 121} (2018) no.~25, 251601},
  \href{http://arxiv.org/abs/1705.03023}{{\tt arXiv:1705.03023 [hep-th]}}.

\bibitem{Garcia-Saenz:2018ifx}
S.~Garcia-Saenz, S.~Renaux-Petel, and J.~Ronayne, ``{Primordial fluctuations
  and non-Gaussianities in sidetracked inflation}'',
  \href{http://dx.doi.org/10.1088/1475-7516/2018/07/057}{{\em JCAP} {\bf 07}
  (2018)  057}, \href{http://arxiv.org/abs/1804.11279}{{\tt arXiv:1804.11279
  [astro-ph.CO]}}.

\bibitem{Bjorkmo:2019aev}
T.~Bjorkmo and M.~D. Marsh, ``{Hyperinflation generalised: from its attractor
  mechanism to its tension with the `swampland conditions'}'',
  \href{http://dx.doi.org/10.1007/JHEP04(2019)172}{{\em JHEP} {\bf 04} (2019)
  172}, \href{http://arxiv.org/abs/1901.08603}{{\tt arXiv:1901.08603
  [hep-th]}}.

\bibitem{Bjorkmo:2019fls}
T.~Bjorkmo, ``{Rapid-Turn Inflationary Attractors}'',
  \href{http://dx.doi.org/10.1103/PhysRevLett.122.251301}{{\em Phys. Rev.
  Lett.} {\bf 122} (2019) no.~25, 251301},
  \href{http://arxiv.org/abs/1902.10529}{{\tt arXiv:1902.10529 [hep-th]}}.

\bibitem{Aragam:2019omo}
V.~Aragam, S.~Paban, and R.~Rosati, ``{Multi-field Inflation in High-Slope
  Potentials}'', \href{http://dx.doi.org/10.1088/1475-7516/2020/04/022}{{\em
  JCAP} {\bf 04} (2020)  022}, \href{http://arxiv.org/abs/1905.07495}{{\tt
  arXiv:1905.07495 [hep-th]}}.

\bibitem{FatI}
D.~Chakraborty, R.~Chiovoloni, O.~Loaiza-Brito, G.~Niz, and I.~Zavala, ``{Fat
  inflatons, large turns and the $\eta$-problem}'',
  \href{http://dx.doi.org/10.1088/1475-7516/2020/01/020}{{\em JCAP} {\bf 01}
  (2020)  020}, \href{http://arxiv.org/abs/1908.09797}{{\tt arXiv:1908.09797
  [hep-th]}}.

\bibitem{DP}
E.~Dimastrogiovanni and M.~Peloso, ``{Stability analysis of chromo-natural
  inflation and possible evasion of Lyth's bound}'',
  \href{http://dx.doi.org/10.1103/PhysRevD.87.103501}{{\em Phys. Rev. D} {\bf
  87} (2013) no.~10, 103501}, \href{http://arxiv.org/abs/1212.5184}{{\tt
  arXiv:1212.5184 [astro-ph.CO]}}.

\bibitem{PPU}
A.~Papageorgiou, M.~Peloso, and C.~Unal, ``{Nonlinear perturbations from
  axion-gauge fields dynamics during inflation}'',
  \href{http://dx.doi.org/10.1088/1475-7516/2019/07/004}{{\em JCAP} {\bf 07}
  (2019)  004}, \href{http://arxiv.org/abs/1904.01488}{{\tt arXiv:1904.01488
  [astro-ph.CO]}}.

\bibitem{CO}
P.~Candelas and X.~de~la Ossa, ``{Moduli Space of {Calabi-Yau} Manifolds}'',
\href{http://dx.doi.org/10.1016/0550-3213(91)90122-E}{{\em Nucl. Phys.} {\bf
  B355} (1991)  455--481}.

\bibitem{BBHL}
K.~Becker, M.~Becker, M.~Haack, and J.~Louis, ``{Supersymmetry breaking and
  alpha-prime corrections to flux induced potentials}'',
  \href{http://dx.doi.org/10.1088/1126-6708/2002/06/060}{{\em JHEP} {\bf 06}
  (2002)  060},
\href{http://arxiv.org/abs/hep-th/0204254}{{\tt arXiv:hep-th/0204254
  [hep-th]}}.

\bibitem{BaBe}
V.~Balasubramanian and P.~Berglund, ``{Stringy corrections to Kahler
  potentials, SUSY breaking, and the cosmological constant problem}'',
  \href{http://dx.doi.org/10.1088/1126-6708/2004/11/085}{{\em JHEP} {\bf 11}
  (2004)  085},
\href{http://arxiv.org/abs/hep-th/0408054}{{\tt arXiv:hep-th/0408054
  [hep-th]}}.

\bibitem{JL1}
H.~Jockers and J.~Louis, ``{The Effective action of D7-branes in N = 1
  Calabi-Yau orientifolds}'',
  \href{http://dx.doi.org/10.1016/j.nuclphysb.2004.11.009}{{\em Nucl. Phys. B}
  {\bf 705} (2005)  167--211}, \href{http://arxiv.org/abs/hep-th/0409098}{{\tt
  arXiv:hep-th/0409098}}.

\bibitem{JL2}
H.~Jockers and J.~Louis, ``{D-terms and F-terms from D7-brane fluxes}'',
  \href{http://dx.doi.org/10.1016/j.nuclphysb.2005.04.011}{{\em Nucl. Phys. B}
  {\bf 718} (2005)  203--246}, \href{http://arxiv.org/abs/hep-th/0502059}{{\tt
  arXiv:hep-th/0502059}}.

\bibitem{GKPW}
T.~W. Grimm, M.~Kerstan, E.~Palti, and T.~Weigand, ``{On Fluxed Instantons and
  Moduli Stabilisation in IIB Orientifolds and F-theory}'',
  \href{http://dx.doi.org/10.1103/PhysRevD.84.066001}{{\em Phys. Rev. D} {\bf
  84} (2011)  066001}, \href{http://arxiv.org/abs/1105.3193}{{\tt
  arXiv:1105.3193 [hep-th]}}.

\bibitem{Gluscevic:2010vv}
V.~Gluscevic and M.~Kamionkowski, ``{Testing Parity-Violating Mechanisms with
  Cosmic Microwave Background Experiments}'',
  \href{http://dx.doi.org/10.1103/PhysRevD.81.123529}{{\em Phys. Rev.} {\bf
  D81} (2010)  123529},
\href{http://arxiv.org/abs/1002.1308}{{\tt arXiv:1002.1308 [astro-ph.CO]}}.

\bibitem{Langlois:2008mn}
D.~Langlois and S.~Renaux-Petel, ``{Perturbations in generalized multi-field
  inflation}'', \href{http://dx.doi.org/10.1088/1475-7516/2008/04/017}{{\em
  JCAP} {\bf 0804} (2008)  017},
\href{http://arxiv.org/abs/0801.1085}{{\tt arXiv:0801.1085 [hep-th]}}.

\bibitem{Weinberg:2008zzc}
S.~Weinberg, {\em Cosmology}.
\newblock Cosmology. OUP Oxford, 2008.
\newblock \url{https://books.google.it/books?id=nqQZdg020fsC}.

\bibitem{WGC}
N.~Arkani-Hamed, L.~Motl, A.~Nicolis, and C.~Vafa, ``{The String landscape,
  black holes and gravity as the weakest force}'',
  \href{http://dx.doi.org/10.1088/1126-6708/2007/06/060}{{\em JHEP} {\bf 06}
  (2007)  060}, \href{http://arxiv.org/abs/hep-th/0601001}{{\tt
  arXiv:hep-th/0601001}}.

\bibitem{FI1}
M.~Cicoli, C.~P. Burgess, and F.~Quevedo, ``{Fibre Inflation: Observable
  Gravity Waves from IIB String Compactifications}'',
  \href{http://dx.doi.org/10.1088/1475-7516/2009/03/013}{{\em JCAP} {\bf 0903}
  (2009)  013},
\href{http://arxiv.org/abs/0808.0691}{{\tt arXiv:0808.0691 [hep-th]}}.

\bibitem{Thesis}
J.~Holland, {\em {}}
\newblock PhD thesis, Swansea U., 2020.

\end{thebibliography}\endgroup

\end{document}